\documentclass{jfm}

\usepackage{amssymb, amsmath, bm, mathtools,gensymb}
\usepackage{subcaption,graphicx}
\usepackage[dvipsnames]{xcolor}
\usepackage{soul}
\usepackage{cleveref}
\usepackage{multirow}
\usepackage{bm,xcolor}

\newcommand\refOne[1]{{\color{black}#1}}
\newcommand\refTwo[1]{{\color{black}#1}}
\newcommand\refThree[1]{{\color{black}#1}}

\newcommand\blue[1]{{\color{black}#1}}

\newcommand{\eps}{\bm{\varepsilon}}
\newcommand{\dd}[1]{{\text{ d}{#1}}}

\newcommand{\cylinderPos}{{\bm{r}}} 	
\newcommand{\sensors}{\bm{s}} 		
\newcommand{\Signal}{\bm{F}} 	
\newcommand{\Measured}{\bm{y}} 		

\newcommand{\CorLen}{\ell} 		

\newcommand{\COMMA}{\, ,}
\newcommand{\PERIOD}{\, .}

\newcommand{\PAR}{\bm{r}} 	
\newcommand{\DES}{\bm{s}} 		
\newcommand{\DATA}{\bm{y}} 		

\newcommand{\PSPACE}{\mathcal{R}}
\newcommand{\YSPACE}{\mathcal{Y}}

\DeclareMathOperator*{\argmax}{argmax}

\newcommand{\movieRotCyl}{Movie 1}
\newcommand{\movieHorizCyl}{Movie 2}
\newcommand{\movieHorizCylStatic}{Movie 3}
\newcommand{\movieDcyl}{Movie 4}

\graphicspath{{./figures/}}

\begin{document}
\setstcolor{red}

\newtheorem{lemma}{Lemma}
\newtheorem{corollary}{Corollary}

\shorttitle{Optimal sensor placement} 
\shortauthor{S. Verma, C. Papadimitriou, N. L{\"u}then, G. Arampatzis, and P. Koumoutsakos} 

\title{Optimal sensor placement for artificial swimmers}

\author
 {
Siddhartha Verma\aff{1,3,4}, Costas Papadimitriou\aff{2},
Nora L{\"u}then\aff{1}, Georgios Arampatzis \aff{1}, and Petros Koumoutsakos\aff{1}
\corresp{\email{petros@ethz.ch}}
  }

\affiliation
{
\aff{1}
Computational Science and Engineering Laboratory, Clausiusstrasse 33, ETH Z{\"u}rich, CH-8092, Switzerland
\aff{2}
Department of Mechanical Engineering, University of Thessaly, Pedion Areos, GR-38334 Volos, Greece
\aff{3}
Department of Ocean and Mechanical Engineering, Florida Atlantic University, Boca Raton, FL 33431, USA
\aff{4}
Harbor Branch Oceanographic Institute, Florida Atlantic University, Fort Pierce, FL 34946, USA
}

\maketitle

\begin{abstract}

Natural swimmers rely for their survival on sensors that gather information from the environment and guide their actions. The spatial organization of these sensors,   such as the visual fish system and lateral line, suggests evolutionary selection, but their optimality remains an open question.
Here, we identify sensor configurations that enable swimmers to maximize the information gathered from their surrounding flow field.  We examine two-dimensional, self-propelled and stationary swimmers that are exposed to disturbances generated by oscillating, rotating and D-shaped cylinders.
We combine simulations of the  Navier-Stokes equations with  Bayesian experimental design to determine the optimal arrangements of shear and pressure sensors that best identify the locations of the disturbance-generating sources. 
We find a marked tendency for shear stress sensors to be located in the head and the tail of the swimmer, while they are absent from the midsection. In turn, we find a high density of pressure sensors in the head along with a uniform distribution along the entire body.  The resulting optimal sensor arrangements resemble neuromast distributions observed in fish and provide evidence for optimality in sensor distribution for natural swimmers.

\end{abstract}

\section{Introduction}

\begin{figure}
        \centering
        \begin{subfigure}[b]{\textwidth}
                \centering
                \includegraphics[width=\textwidth]{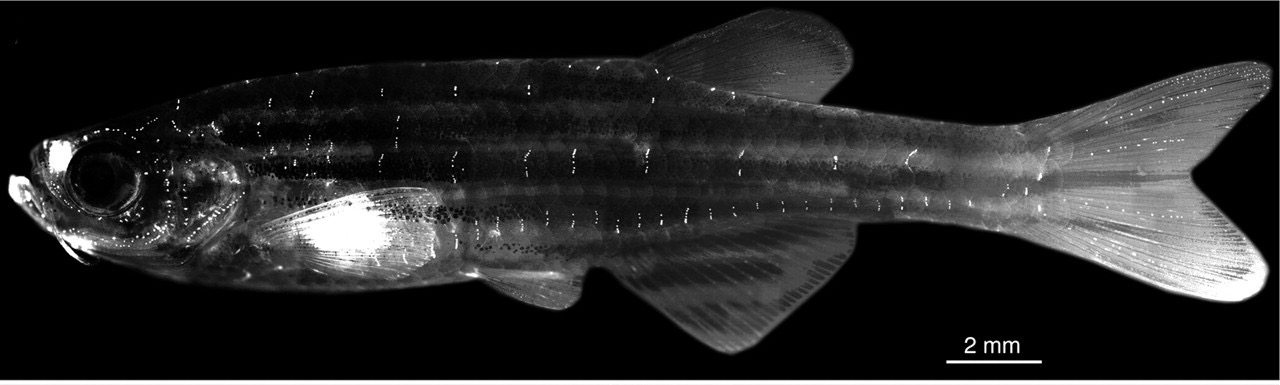}
		\subcaption{}
                \label{fig:lateralLineSystem}
        \end{subfigure} \\
        \begin{subfigure}[b]{0.65\textwidth}
                \centering
                \includegraphics[width=\textwidth]{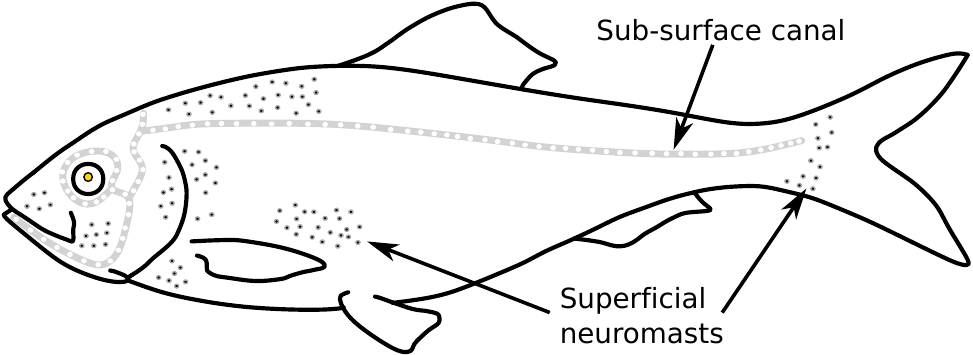}
		\subcaption{}
                \label{fig:lateralLineSketch}
        \end{subfigure} \qquad \qquad
        \begin{subfigure}[b]{0.2\textwidth}
                \centering
                \includegraphics[width=\textwidth]{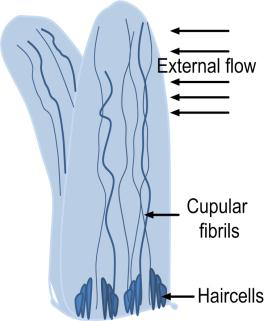}
                \subcaption{}
                \label{fig:neuromastsSketch}
        \end{subfigure}
	\caption{(\subref{fig:lateralLineSystem}) The lateral line in juvenile zebrafish, with neuromasts visible as bright dots on the body surface (adapted with permission from \citet{Sapede2002}). We observe a high density of neuromasts in the head and the tail, with sparser distribution along the midsection. (\subref{fig:lateralLineSketch}) A schematic representation of the distribution of mechanoreceptors along the fish body. (\subref{fig:neuromastsSketch}) The  neuromasts bend in response to  flow, which generates \blue{a} neuronal response  \blue{by} sensory cells located at the base (adapted with permission from \citet{Kottapalli2013}).}
\label{fig:fishNeuromasts}
\end{figure}

The capability of aquatic animals to accurately perceive their environment plays a crucial role in their  survival. 
Many fish species employ specialized organs to obtain visual, olfactory, and tactile cues from their environment \blue{ which} often complement each other.  Predator-detection by fish using visual or olfactory cues~\citep{Hara1975,Ladich2003,Valentincic2004} is crucial for providing early-warning\blue{, since} mechanical disturbances may be imperceptible \blue{at large distances}. On the other hand, sensory organs specialized for detecting mechanical disturbances~\citep{Schwartz1974} take precedence when fish operate in deep or turbid waters, where visual and other sensory mechanisms may become ineffective. In these situations, the burden of collecting sensory information falls primarily on the `lateral line' organ in fish~\citep{Dijkgraaf1963,Kroese1992,Coombs1996,Coombs2005,Bleckmann2009}. These organs are comprised of hair-like mechanoreceptors called neuromasts (Figure~\ref{fig:fishNeuromasts}), which generate neuronal impulses when deflected by either the flow shear (superficial neuromasts - \cite{engelmann2000neurobiology}) or non-zero pressure gradients (sub-surface `canal' neuromasts - \cite{Bleckmann2009}). An array of such sensors allows fish to discern both \blue{the} direction and speed of disturbances generated in the\blue{ir} surrounding flow~\citep{Chambers2014,Asadnia2015}.

\blue{These} flow \blue{sensors}  are  distributed in distinctive patterns on the body, with the \blue{canal neuromasts distributed evenly along the midline from head to tail~\citep{Ristroph2015},}  and \blue{superficial} neuromasts found in dense clusters near the head \blue{and tail,}  with a sparser distribution \blue{along the midsection (Figure~\ref{fig:fishNeuromasts})}. The fact that they are not distributed uniformly over the body, as well as differences \blue{in distribution} among species inhabiting different hydrodynamic environments~\citep{engelmann2000neurobiology, Bleckmann2009, atema1988sensory}, suggest that neuromast distribution may be optimized for characterizing hydrodynamic disturbances.

Experimental studies have demonstrated that a well-functioning lateral line is crucial for a range of routine behaviour, such as schooling~\citep{Pitcher1976,Partridge1980}, predator evasion~\citep{Blaxter1989}, prey detection/capture~\citep{Hoekstra1985}, reproduction~\citep{Satou1994}, rheotaxis~\citep{Dijkgraaf1963,Kanter2003}, obstacle avoidance~\citep{Hassan1989}, and station-keeping by countering the effects of unsteady gusts~\citep{Sutterlin1975}. Disrupting the normal functioning of the lateral line, either via chemical or mechanical means, hinders fish's ability to perform these tasks effectively. \citet{Liao2006} demonstrated that disabling the lateral line system influences fish's ability to harness energy from unsteady flows. The sensory system also plays a vital role in `hydrodynamic imaging', where fish devoid of visual cues swim past walls and unknown objects repeatedly to form a hydrodynamic `map' of their surroundings~\citep{Hassan1989,Coombs1999,Montgomery2001,Coombs2003}. Certain species such as the Blind Cave Fish, which have evolved degenerated sight, rely heavily on this technique for navigation, and for inferring the shape and size of unfamiliar objects~\citep{vonCampenhausen1981,Windsor2008,dePerera2004}.

The lateral line system has inspired the design of artificial sensory arrays, given their potential to transform underwater navigation of robotic vehicles~\citep{Yang2006,Yang2010,Kottapalli2012,Jezov2012,Kruusmaa2014,Asadnia2015,Triantafyllou2016,Strokina2016,Kottapalli2018,Yen2018}. Such mechanoreceptors would be a vital addition to the already available suite of visual and acoustic sensors, with the added advantage of low energy-consumption\blue{,} since they operate via passive mechanical deformation. \blue{These}  vibration-detecting sensors \blue{would}  be crucial for navigation, detection, and tracking in low-light conditions, or in scenarios where the use of onboard lights or sonar is undesirable, either for maintaining stealth, or for minimally intrusive observation of animals. Current prototypes of such artificial sensors are based on arrays of pressure transducers~\citep{Fernandez2011,Venturelli2012,Xu2017}, and  mechanically deforming hair-like structures~\citep{Yang2006,Tao2012,Abdulsadda2013,Dagamseh2013,deVries2015,Triantafyllou2016}.

The importance of the lateral line as an essential sensory organ in fish, and its immense potential for driving the bio-inspired design of artificial sensors, has stimulated numerous experimental and model-based studies. The structure and function of these sensory arrays has been investigated via biological experiments, to characterize their response to pressure differences and object-induced vibrations in water~\citep{Gray1984,Denton1988,Kroese1992,Coombs1996,Blake2006}. Experiments using artificial fish models have tried to emulate these biological studies, using pressure-transducers and hair-like sensors to characterize the frequency and range of oscillating spheres \citep{Montgomery1998}, and Karman vortex streets \citep{Venturelli2012}. Moreover, there have been a number of mathematical model-based studies, that have combined potential-flow solutions with simplified representations of fish-swimming to study the functioning of the lateral line~\citep{Hassan1992,Franosch2009,Bouffanais2011,Ren2012,Colvert2016}. A few of these studies have attempted to infer the optimal arrangement of sensors on rigid objects exposed to various flow conditions. \citet{Colvert2016} determined the optimal placement of a single sensor-pair on an elliptical body, moving at different orientations in uniform flow. \citet{Ahrari2017} used simplified analytical representations to determine optimal sensor-arrangement and -orientation on a rigid hydrofoil, which could best characterize a dipole source with six degrees of freedom in three dimensions.

While model-based studies provide important insight regarding sensing, they suffer from certain drawbacks owing to simplified hydrodynamics, and simplistic representations of fish-swimming (e.g., ellipses and rigid airfoils). Neglecting the effects of viscosity in potential-flow based studies is a notable disadvantage, especially when considering larvae swimming at relatively low Reynolds numbers \blue{(\textit{Re})}. Moreover, viscous effects play \blue{a substantial}  role in the operation of the lateral line \citep{Triantafyllou2016}, given that superficial neuromasts are immersed in the fish's boundary layer, and  canal neuromasts encounter low $Re$ flow inside constricted channels. The Reynolds number that  animals operate at can also have a considerable impact on the functioning of the lateral line \citep{Webb2014}, which cannot be accounted-for via inviscid assumptions. The importance of viscous effects has also been demonstrated by \citet{Rapo2009}, who studied the impact of an oscillating sphere on the boundary layer of a vibrating flat plate, albeit using analytical simplifications to circumvent the high computational cost of three-dimensional numerical simulations. \refOne{Recent studies using two-dimensional viscous computations have attempted to classify wake patterns behind an oscillating airfoil using Artificial Neural Networks~\citep{Colvert2018,Alsalman2018}. Using flow sensors placed in the wake of the airfoil, they determine that both the spatial distribution of the sensors as well as the flow variable being measured influence the accuracy for predicting wake characteristics.} Here, we investigate the role of hydrodynamics in determining the sensor-distribution observed in fish, using two-dimensional Navier-Stokes simulations of self-propelled swimmers to overcome  the limitations mentioned above. We determine the optimal spatial distribution of sensors via Bayesian optimal experimental design, and we find that the resulting patterns are closely related to sensory layouts found in natural swimmers.

\section{Methods}
\label{sec:methods}

The present study relies on two-dimensional simulations of  a self-propelled swimmer possessing shear stress and pressure gradient sensors  on its surface. The swimmer is exposed to disturbances generated by cylinders located at various positions in the environment. The  sensor locations are identified by formulating a Bayesian optimal experimental design with the goal of maximizing  the information gain  of the swimmer in its environment.

\subsection{Numerical methods}
\label{sec:numMeth}

We conduct two-dimensional simulations of viscous flows past multiple bodies by discretizing the  vorticity form of the incompressible Navier-Stokes equations,
\begin{equation}
\dfrac{\partial \omega}{\partial t} + (\bm{u}\cdot\nabla)\omega = \nu\nabla^2\omega + \lambda\nabla\times\left(\chi\left(\bm{u}_s-\bm{u}\right)\right) \COMMA
\label{eq:penalNSvort}
\end{equation}
where $\bm{u}$ is the flow-velocity, and $\omega = \nabla \times {\bm{u}}$ is the vorticity. The penalty term, $\lambda\nabla\times\left(\chi\left(\bm{u}_s-\bm{u}\right)\right)$ models the interaction of objects with the surrounding fluid (\citet{coquerelle2008vortex}), where $0<\chi\le 1$ indicates the solid body. \blue{$\lambda$} is the penalization parameter, and $\bm{u}_s$ represents the combined translational, rotational, and deformational velocity of the solid object. The equations are discretized using remeshed vortex methods \citep{Koumoutsakos1995} and wavelet adapted grids \citep{Rossinelli2015}, and the  penalty term is integrated via the fully implicit backward Euler method. Additional details for the computational methods may be found in \citet{Gazzola2011} and \citet{Rossinelli2015}. 
\blue{The simulation domain is a unit square, with an effective resolution of $4096^2$ grid points.} The fish length is  $L= 0.2$ \blue{units}, with approximately 800 grid points along its mid-line.

\subsection{Swimmer shape and kinematics}
\label{sec:shapeAndKinematics}
\begin{figure}
        \centering
        \begin{subfigure}[b]{0.49\textwidth}
                \centering
                \includegraphics[width=\textwidth]{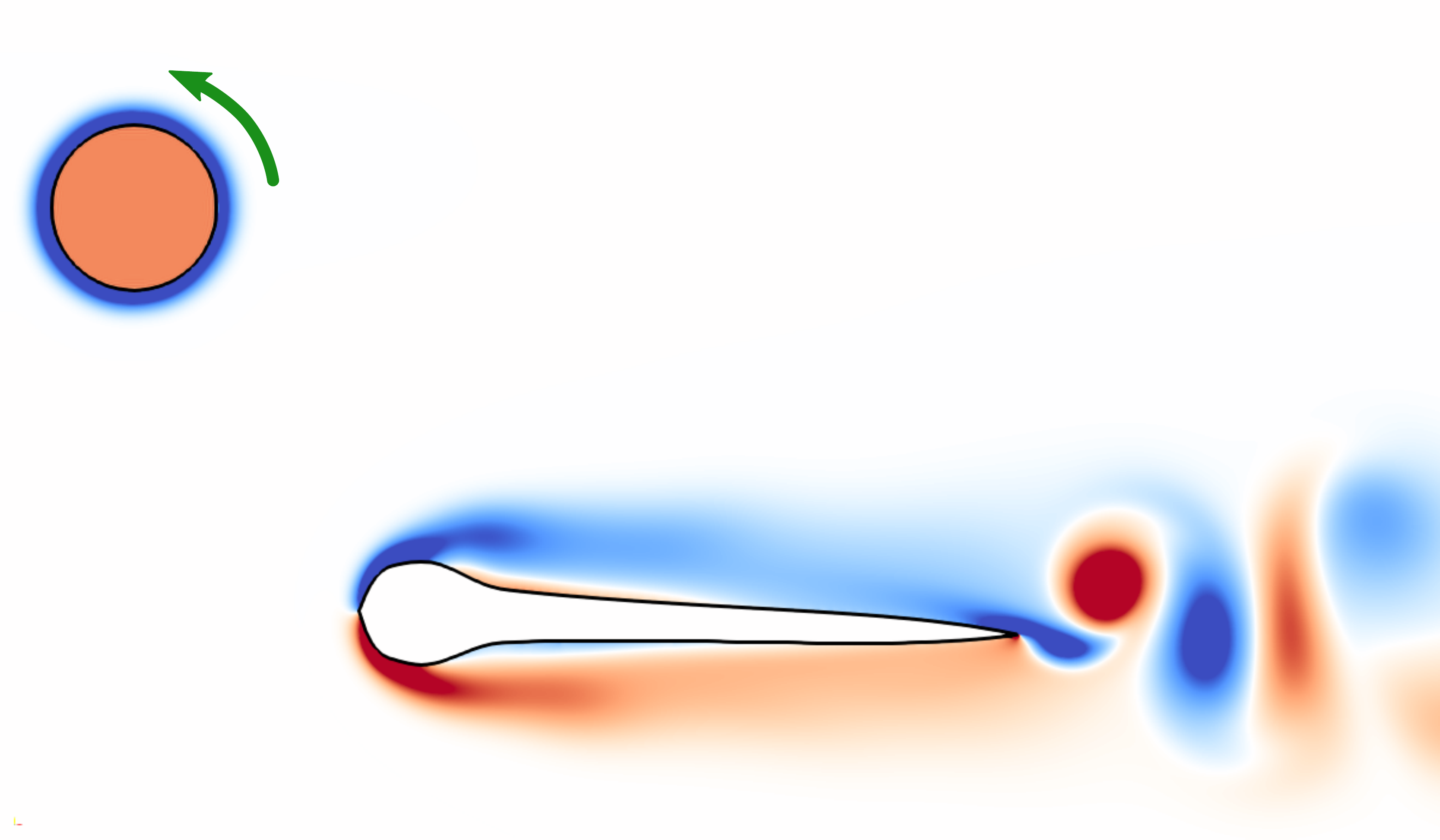}
                \subcaption{}
                \label{fig:larvaShape}
        \end{subfigure}
        \begin{subfigure}[b]{0.49\textwidth}
                \centering
                \includegraphics[width=\textwidth]{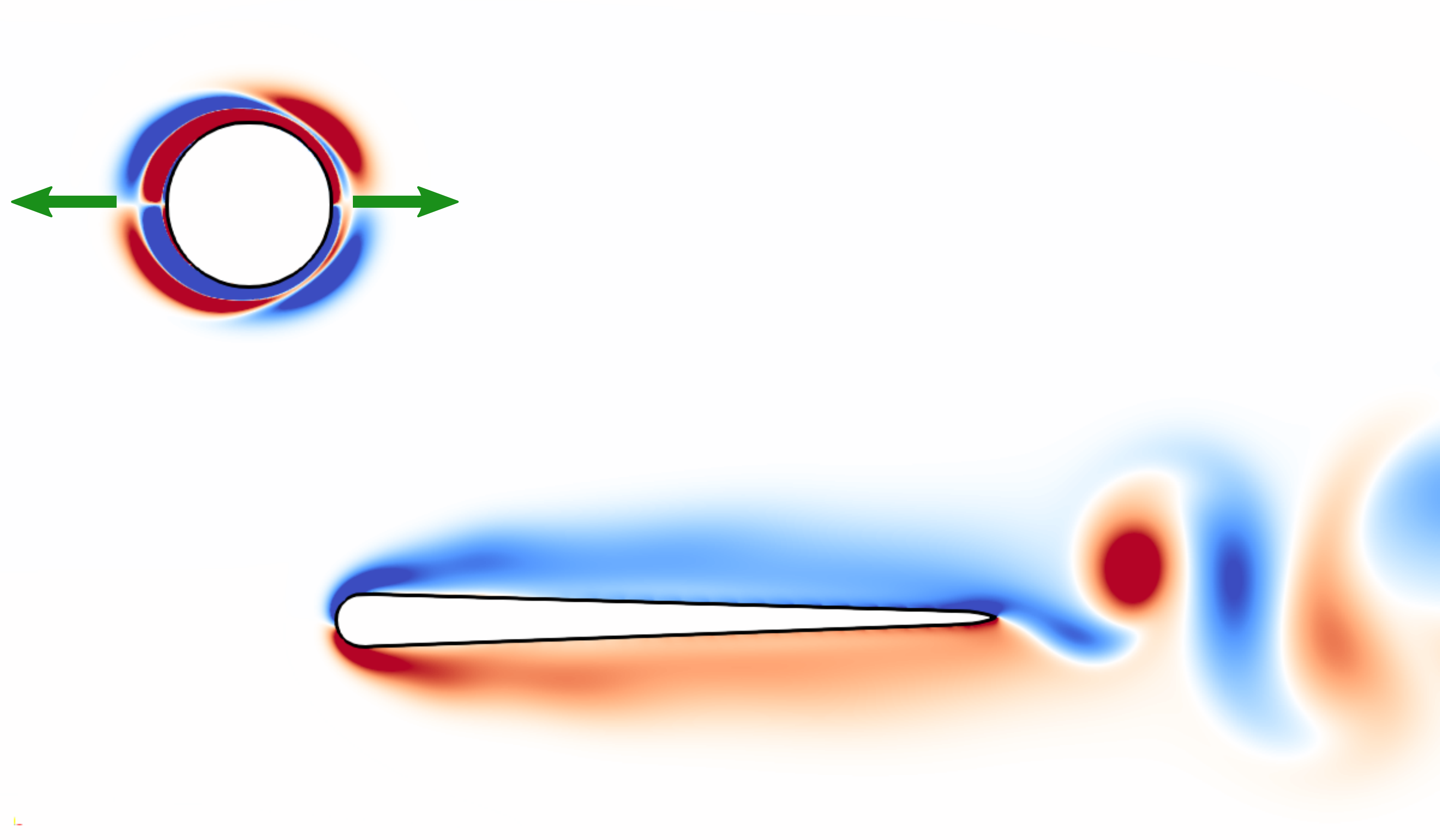}
                \subcaption{}
                \label{fig:adultShape}
        \end{subfigure}
		\caption{(\subref{fig:larvaShape}) A larva-shaped swimmer detecting disturbances generated by a rotating cylinder (angular velocity = \blue{10 rotations/s}) . Regions \blue{with}  positive vorticity are coloured in red, and those with negative vorticity are coloured in blue. (\subref{fig:adultShape}) An adult-shaped swimmer detecting an oscillating cylinder (\blue{amplitude=$0.075L$,} frequency = $10Hz$). Animations for these two cases are shown in \blue{supplementary} \movieRotCyl{} and \movieHorizCyl{}.}
\label{fig:showShape}
\end{figure}

We consider two distinct scenarios for the swimmer behaviour to identify the optimal distribution of sensors: one where external disturbances are detected by a static fish-shaped body, and the other involving a self-propelled swimmer. Furthermore, we examine the influence of body geometry on optimal sensor distribution by considering two  shapes for the swimmers modelled after  zebrafish in their larval  and adult stages. The larva shape, shown in Figure~\ref{fig:larvaShape}, is based on silhouettes extracted from experiments, whereas the adult fish is modelled using a geometric combination of circular arcs, lines, and parabolic sections (Figure~\ref{fig:adultShape}) \citep{Gazzola:2012}. Details regarding shape parametrization for both cases are provided in the Appendix (Eqs.~\ref{eq:adultShape} and~\ref{eq:larvaShape}).
The swimmers propel themselves by imposing a sinusoidal wave travelling along the body. Details of the swimming kinematics are also provided in Appendix~\ref{app:shapeKinematics}.


\subsection{Disturbance-generation and detection}
\begin{figure}
	\centerline{\includegraphics[width=0.75\textwidth]{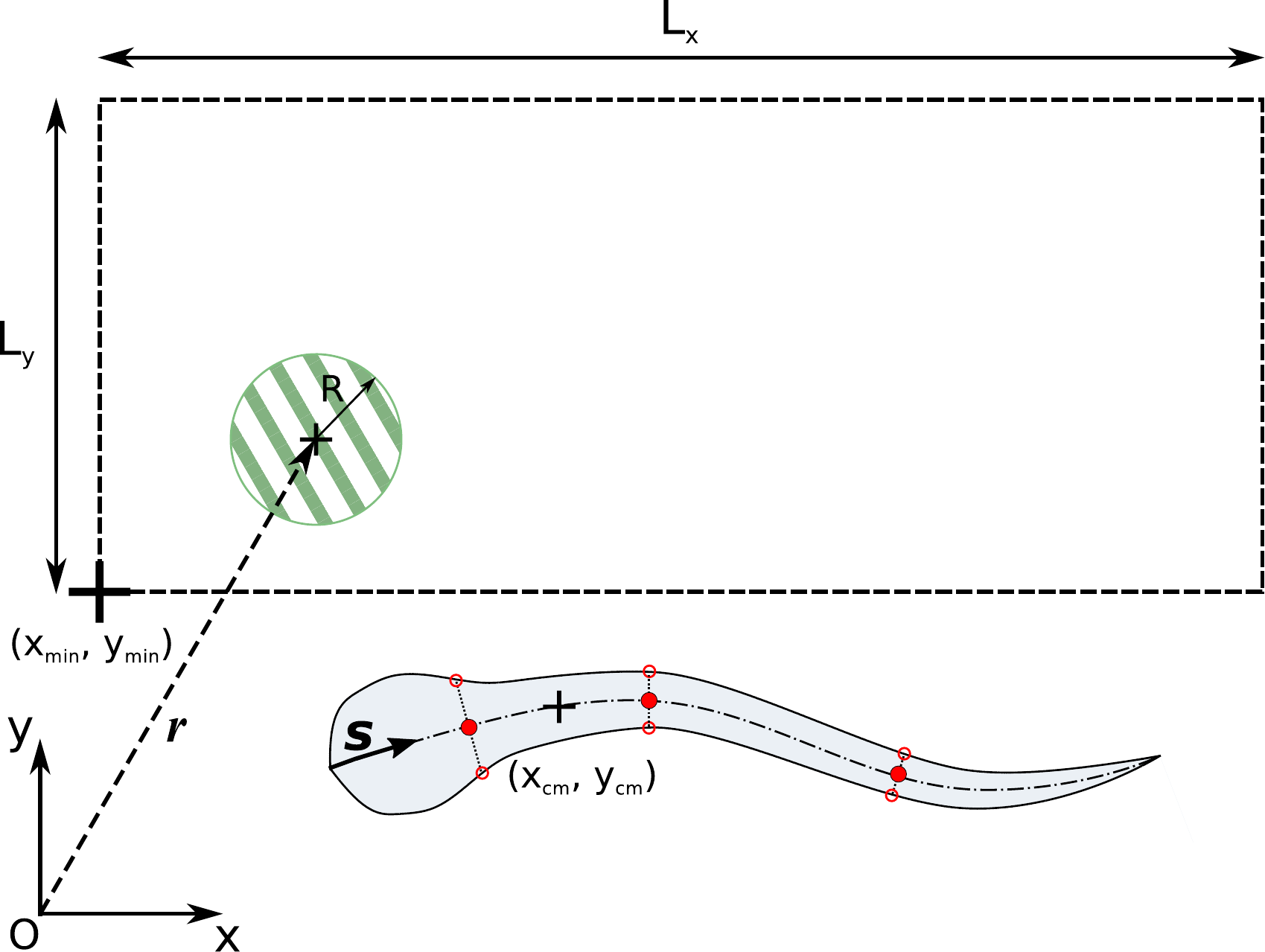}}
	\caption{Setup used for determining the optimal sensor distribution on a fish-like body swimming past a cylinder located within the rectangular area, which is referred to as the `prior-region'.  The sensor-placement algorithm attempts to find the best arrangement of sensors that allows the swimmer to identify the correct cylinder position with minimal uncertainty.}
 \label{fig:priorRegion}
\end{figure}

The sensory cues detected by \blue{the} rigid and swimming bodies \blue{described in section~\ref{sec:shapeAndKinematics}} are generated using oscillating and rotating cylinders of diameter $D=0.25L$ (Figure~\ref{fig:showShape}), and  a D-shaped half cylinder of diameter $0.5L$. The amplitude and frequency of the horizontally oscillating cylinders are set to $A_{cyl} = 0.075L$ and $f_{cyl} = 10Hz$, whereas the angular velocity of the rotating cylinders is set to $20\pi \ rad/s$ (10 rotations/s). 
The cylinders are placed at various locations within a prescribed region in the computational domain (in  the `prior-region'), as shown in Figure~\ref{fig:priorRegion}.

We distinguish two types of sensors on the swimmer body. Shear stress sensors estimate the local shear stress by measuring the tangential flow velocity in the reference-frame of the swimmer at 2 grid cells away from the body, corresponding to a physical distance of $0.0024L$.
These  \blue{sensors} are analogous to superficial neuromasts in fish that protrude into the boundary layer and measure tangential velocity \citep{Kroese1992,Bleckmann2009,Asadnia2015}. In addition, we consider pressure gradient sensors that correspond to canal neuromasts observed in natural swimmers. We compute pressure gradient along the swimmers' surface by fitting a least-squares cubic spline to surface pressure, in order to minimize derivative noise. \refOne{We note that the sensor-measurements are scalar quantities, since both the shear stress and pressure gradient are projected along the surface tangent vector at each measurement location on the body. This is done so that the measured scalar quantities are a close representation of the flow-induced forces that deflect the hair-like sensory structures in real fish.}

In the case of self-propelled swimmers, measurements are taken towards the end of a coasting phase to allow self-generated disturbances to subside sufficiently, and are averaged over a small time window from $15.750T$ to $15.875T$. For  motionless larvae, the disturbance-sources start moving at $0$s, and time-averaging of the recorded data is done between $0.95$s and $1.0$s. This allows transients from the initial cylinder start-up to dissipate sufficiently. Time averaging for the D-cylinder simulations is done from $18$s to $20$s, which allows adequate time for vortex shedding to exhibit a periodically repeating pattern. These measurements are then used to determine the optimal arrangement of sensors on the swimmer body, via the Bayesian optimal experimental design algorithm described in section~\ref{sec:optimalSensor}.

\subsection{Bayesian optimal sensor placement}
\label{sec:optimalSensor}

\subsubsection{Bayesian estimation of disturbance location}
We consider a disturbance-generating source (for example an oscillating or rotating cylinder) located at coordinates $\cylinderPos=\left(x, y\right)$ in the region shown in Figure~\ref{fig:priorRegion}. The uncertainty in the values of the coordinates of the cylinder is quantified by a probability distribution that is updated based on measurements collected on the surface of the swimmers.  The cylinder location can be detected provided that disturbances induced by the cylinder to the surrounding fluid  are detected by sensors located on the swimmer surface. The problem of optimal placement implies that we identify the configuration of  sensors that can provide the best estimate for the coordinates of the cylinder ($\cylinderPos$). We assume that the sensor locations  are placed  symmetrically on both  sides of the two-dimensional swimmer and they are described by a vector $\bm{s}\in R^{n}$, that is the  mid-line coordinate of each sensor pair, with values in $[0,L]$(see Figure 3).  The shear stress  or the pressure gradient are measured on the surface points corresponding to the positions  $\sensors$, and are listed in a vector $\bm{y}\in R^{2n}$.

We denote as $\Signal( \cylinderPos; \sensors)$ the predictions of shear stress /pressure gradient at sensor locations $\sensors$, obtained by solving the Navier-Stokes equations with a disturbance-generating source located at $\cylinderPos$.
Moreover, we assume that we have prior knowledge about the parameter $\cylinderPos$, encoded in a \emph{prior} probability distribution $p(\cylinderPos)$.  

After observing the measurements $\Measured$ from sensors $\sensors$, we use Bayesian inference to update our prior belief for the plausible values of parameter $\cylinderPos$, by identifying the \emph{posterior} probability distribution $p(\cylinderPos | \Measured ,\sensors)$. 
 Following Bayes' rule, the posterior distribution $p(\cylinderPos | \Measured ,\sensors)$ of the model parameters is proportional to the product of the prior distribution $p(\cylinderPos)$ and the \emph{likelihood} $p(\Measured | \cylinderPos,\sensors)$. The likelihood function represents the probability that a particular measurement $\Measured$ for a given sensor arrangement $\sensors$ originates from the disturbance-source located at $\cylinderPos$. We assume a prediction error, $\eps(\sensors)$, as the difference between the measurements $\Measured$ and the predictions $\Signal( \cylinderPos; \sensors)$ such that:
\begin{equation}
	\Measured = \Signal( \cylinderPos; \sensors) + \eps(\sensors) \PERIOD
\label{eq:measurement}
\end{equation}
The prediction error term ($\eps(\sensors)$) represents errors that can be attributed to measurement- and  model-errors, as well as  numerical errors due to spatio-temporal discretization of the Navier-Stokes equations. Following the maximum entropy criteria the prediction error $\eps(\sensors)$ follows a multivariate Gaussian distribution $\mathcal{N}(0, \Sigma(\sensors))$ with zero mean and  covariance matrix $\Sigma(\sensors) \in R^{2n \times 2n}$. The likelihood function $p(\Measured \vert \cylinderPos,\sensors)$ is then expressed as:
\begin{align}
p\left(\Measured \vert \cylinderPos, \sensors\right) &= 
\dfrac{1}{\sqrt{(2\pi)^{2n} \det(\bm{\Sigma}(\sensors))}} \exp \left(-\dfrac{1}{2} 
	\left(\Measured - \Signal( \cylinderPos; \sensors) \right)^T 
	\bm{\Sigma}^{-1}(\sensors) 
	\left(\Measured - \Signal( \cylinderPos; \sensors)  \right) \right) \PERIOD
\label{eq:likelihood}
\end{align}

\subsubsection{Optimal sensor placement based on information gain}

The goal of the \emph{optimal sensor placement} problem is to find the locations $\sensors$ of the sensors such that the data measured in these locations are most informative for estimating the position $\cylinderPos$ of the disturbance. A measure of information gain is provided by the Kullback-Leibler (KL) divergence between the prior and the posterior distribution.  We postulate that the optimal sensor configuration maximizes a utility function that represents the information gain, or equivalently, the Kullback-Leibler divergence defined as:
\begin{equation}
 u(\sensors,\Measured) := \int_{\PSPACE} p( \cylinderPos | \Measured,\sensors) \; \ln \frac{p(\cylinderPos | \Measured,\sensors)}{p(\cylinderPos)} \dd{\cylinderPos} \PERIOD
\label{eq:utilityFunction}
\end{equation}
We note that in the experimental design phase, the measurements $\Measured$ are not available.  Thus, the prediction error model (Eq.~\ref{eq:measurement}) is used to generate measurements for given model parameter values $\cylinderPos$ and sensor configuration $\sensors$. We identify the best sensor arrangement by maximizing a utility function, defined as the expected value of the Kullback-Leibler divergence over all possible values of the measurements simulated by Eq.~\ref{eq:measurement}~\citep{Ryan2003}:
\begin{equation}
\begin{split}
 U(\DES) := \mathbb{E}_{\Measured | \sensors} \big [ u(\sensors,\Measured) \big ] &= \int_{\YSPACE} u(\sensors,\Measured)\; p(\Measured | \sensors) \dd{\Measured}  \\
&= \int_{\YSPACE}  \int_{\PSPACE} p( \cylinderPos | \Measured,\sensors) \; \ln \frac{p(\cylinderPos | \Measured,\sensors)}{p(\cylinderPos)} \; p(\Measured | \sensors) \dd{\cylinderPos} \dd{\Measured} \PERIOD
 \label{eq:utility}
\end{split}
\end{equation}
%
The expected utility function involves a double integral over the parameter space $\cylinderPos$ and over the measured data $\Measured$. An efficient estimator of this double integral using sampling techniques is provided by \citet{huan2013simulation}. A similar estimator is used in the present work,
\begin{equation}
\hat U(\sensors) = \frac{1}{N_{\DATA}} \sum_{j=1}^{N_{\DATA}} \sum_{i=1}^{N_{\cylinderPos}}  w_i p(\cylinderPos^{(i)})  \left [   \ln p(\Measured^{(i,j)} | \cylinderPos^{(i)},\sensors) - \ln \left( \sum_{k=1}^{N_{\cylinderPos}} w_k  p(\cylinderPos^{(k)}) p(\Measured^{(i,j)} | \cylinderPos^{(k)},\sensors)  \right ) \right ] \PERIOD
\label{eq:estimator}
\end{equation}
A detailed derivation and discussion of the estimator is provided in Appendix~\ref{sec:utilityDerivation}. Our estimator employs a quadrature technique to evaluate the integral over the two-dimensional parameter space $\cylinderPos$. In Eq.~\ref{eq:estimator}, $\cylinderPos^{(i)}$ and $w_i$ denote $N_{\cylinderPos}$ quadrature points and corresponding weights related to discretization of the two-dimensional prior-region \blue{(Figure~\ref{fig:priorRegion})}. A total of $N_{\cylinderPos}$ distinct Navier-Stokes simulations are conducted, with a cylinder positioned at \blue{various}  discrete points $\cylinderPos^{(i)}$, and the quadrature is evaluated using the trapezoidal rule. Based on the prediction error defined in Eq.~\ref{eq:measurement}, the measured data $\Measured^{(i,j)}$ in Eq.~\ref{eq:estimator} are given by,
\begin{equation}
	\Measured^{(i,j)} = \Signal( \cylinderPos^{(i)}; \sensors) + \eps^{(j)} \COMMA
	\label{eq:measuredDiscrete}
\end{equation}
where $\eps^{(j)},\, \blue{\text{ with }} j=1,\ldots,N_{\Measured}$,  are vectors sampled from the distribution $\mathcal{N}(0, \Sigma(\sensors))$. $N_{\Measured}$ is set to $100$ in the current work\blue{, which results in a smoother estimate of $\hat U(\sensors)$ in Eq.~\ref{eq:estimator}.}

We note that the computational effort for evaluating \blue{$\hat U(\sensors)$ in Eq.~\ref{eq:estimator}}  depends primarily on the number of Navier-Stokes simulations, $N_{\cylinderPos}$, which are required to evaluate $\Signal(\cylinderPos^{(i)}, \sensors)$ for different disturbance locations $\cylinderPos^{(i)}$, and \blue{subsequently to} determine $\Measured^{(i,j)}$ using Eq.~\ref{eq:measuredDiscrete}. The computational burden does not depend on the number of measured samples $N_{\Measured}$, since there are no additional time consuming simulations involved in generating $\eps^{(j)}$. Thus the computational effort scales linearly with the number $N_r$ of model parameter points $\cylinderPos^{(i)}$. 

We assume that the prior distribution $p(\cylinderPos)$ for the location of the disturbance source is uniform over the prior-region shown in Figure~\ref{fig:priorRegion}, i.e., the probability of finding the source is constant for all locations. Moreover, the only available information we have is a description of the prior-region where the disturbance may be found. Using Bayes' theorem, and the fact that the prior distribution is uniform, we can assert that the posterior distribution of a disturbance location $\cylinderPos$, $p(\cylinderPos \vert \Measured,\sensors)$, is proportional to the likelihood function $p(\Measured \vert \cylinderPos, \sensors)$.

The covariance matrix $\Sigma(\sensors)$ depends primarily on the sensor positions $\sensors$, and is diagonal if the errors at the given sensor positions are independent of each other. In the current work, the prediction errors are assumed to be correlated for measurements collected on the same side of the swimmer (i.e., left- or right-lateral surfaces), and decorrelated if the measurements originate from opposite sides. An exponentially decaying correlation is assumed for the covariance matrix,
\begin{equation}
\Sigma_{ij}(\sensors) =
\begin{cases}
\sigma^2 \exp   \left( - \frac{ \| \bm{x}(s_i) - \bm{x}(s_j) \| }{\CorLen}\right),  & \quad \textrm{ if } 1\leq i,j \leq n, \\
\Sigma_{i-n,j-n}(\sensors),										& \quad \textrm{ if } n < i,j \leq 2n, \\
0														& \quad \textrm{ otherwise,}
\end{cases}
\label{eq:covarianceMatrix}
\end{equation}
where $\bm{x}(s_i)$ corresponds to the coordinates of the $i$-th sensor on the right lateral surface of the swimmer, $\CorLen>0$ is the prescribed correlation length, and $\sigma$ is the correlation strength. For all the simulations described in this work, the correlation length is set to be $\CorLen=0.01L$. The correlation strength $\sigma$ is a fixed percentage ($30\%$) of the mean sensor-measurement, which is computed over all available instances of $\cylinderPos$ and at all points discretizing the swimmer skin. This form of the correlation error reduces the information-gain when sensors are placed too close together \citep{papadimitriou2012effect, simoen2013prediction}, and prevents excessive clustering of sensors within confined neighbourhoods. 

Finally, we provide an intuitive interpretation of how Eq.~\ref{eq:estimator} relates to information gain. Let us assume that a particular set of sensors is able to characterize the disturbance sources quite effectively. Moreover, we assume that the measurement $\Measured^{(i,j)}$ has been generated by a disturbance located at $\cylinderPos^{(i)}$. This implies that the posterior $p(\cylinderPos \vert \Measured^{(i,j)},\sensors)$, which indicates the probability that a particular disturbance source $\cylinderPos$ has generated the measurement measurement $\Measured^{(i,j)}$, is peaked and centered around the true source location $\cylinderPos^{(i)}$. Since the prior distribution is uniform, the likelihood $p(\Measured^{(i,j)} \vert \cylinderPos,\sensors)$ is proportional to the posterior, and is also peaked and centered around $\cylinderPos^{(i)}$. Thus, the first term in Eq.~\ref{eq:estimator} is large, whereas most of the terms in the second sum are close to zero, since the probability of measurement $\Measured^{(i,j)}$ originating from source $\cylinderPos^{(k)}$ is small due to the peaked nature of the posterior (except for $k=i$). In this case, the expected utility value computed using Eq.~\ref{eq:estimator} is large. On the other hand, a poor sensor arrangement which cannot characterize source positions well, yields flatter likelihood and posterior distributions due to high uncertainty. Thus, different source positions yield similar measurements at the selected sensors, which makes the second sum in Eq.~\ref{eq:estimator} larger (non-zero $p(\Measured^{(i,j)} \vert \cylinderPos^{(k)},\sensors)$ even for $k\ne i$), thereby reducing the utility value.

\subsubsection{Optimization of the expected utility function}
\label{sec:sequentialOpt}

The optimal sensor arrangement is obtained by maximizing the expected utility estimator $\hat U(\sensors)$ described in Eq.~\ref{eq:estimator}. However, optimal sensor placement problems are characterized by a relatively large number of multiple local optima. Heuristic approaches, such as the sequential sensor placement algorithm described by \citet{Papadimitriou2004}, have been demonstrated to be effective alternatives. In this approach, the optimization is carried out iteratively, one sensor at a time. First, $\hat U(\sensors)$ is \blue{computed}  for a single sensor\blue{-pair} $\sensors=s_1$, and the optimal solution $s_1^\star$ is obtained by identifying the maximum in $\hat U(\sensors)$. Then, $\hat U(\sensors)$ is recomputed with $\sensors=(s_1^\star,s_2)$, and it is optimized with respect to the second sensor\blue{-pair,} resulting in an optimal solution $s_2^\star$. We can generalize this procedure for all subsequent sensors, by defining $\hat{U}_i(s)  = \hat{U} (s_1^\star,\ldots,s_{i-1}^\star,s)$. The optimal solution for the $i$-th sensor is given as,
\begin{equation}
s_i^\star = \argmax_{s}  \;  \hat{U}_i (s) \qquad \textrm{and} \qquad    \hat{U}_i^\star =   \max_{s} \; \hat{U}_i (s) \PERIOD
\label{eq:optimal:sensor}
\end{equation}
We note that the scalar variable $s$ denotes the position of a single sensor\blue{-pair}, whereas the vector $\sensors$ holds the position of all sensor\blue{-pairs} along the swimmer's midline. 

The sequential placement procedure is carried out for a number of sensors, $N_s$, and it terminates when the last sensor in the optimal configuration is identified $\sensors^\star=(s_1^\star,\ldots,s_{N_s}^\star)$. \citet{Papadimitriou2004} has demonstrated that the heuristic sequential sensor placement algorithm provides a sufficiently accurate approximation of the global optimum. Moreover, using the sequential optimization approach, $N_s$ one-dimensional problems have to be solved, instead of one $N_s$-dimensional problem. We solve each one-dimensional problem of identifying the maximum of $\hat{U}_i$ via a grid search, where the swimmer midline is discretized using the points $\{ \,  k \Delta s, \; k=0,\ldots,N_g \, \} $, with $\Delta s=L/N_g$ and $N_g=1000$. Thus, for each iteration of sequential optimization, the utility estimator in Eq.~\ref{eq:estimator} has to be evaluated $N_g+1$ times.

We remark that the Bayesian optimal design procedure is  computationally demanding, as it entails model simulations for several different sensor configurations $\sensors$. To minimize the relevant computational cost, we run $N_{\cylinderPos}$ distinct Navier-Stokes simulations for all disturbance locations $\cylinderPos^{(i)}$ ($i=1\ldots,N_r$), and store the shear stress  \blue{and pressure gradient} at all available dicretization points along the swimmer skin offline. This allows us to reuse simulation data for a particular disturbance-source, without having to re-run Navier-Stokes simulations for different sensor configurations. We note that the skin discretization may not correspond to the $N_g$ points used for computing $\hat{U}_i^\star$. Thus, the output quantities of interest are averaged at appropriate locations along the swimmer surface, over a small neighbourhood of size $0.01L$.

\section{Results}
\label{sec:results}

We first examine the optimal arrangement of \blue{shear stress and pressure gradient} sensors on motionless larva in the presence of oscillating,  rotating  and D-shaped cylinders. We then consider self-propelled swimmers, which are exposed to cylinder-generated disturbances. 

\subsection{Stationary swimmer in the vicinity of oscillating/rotating cylinders}

\begin{figure}
        \centering
        \begin{subfigure}[b]{0.32\textwidth}
                \centering
                \includegraphics[width=\textwidth]{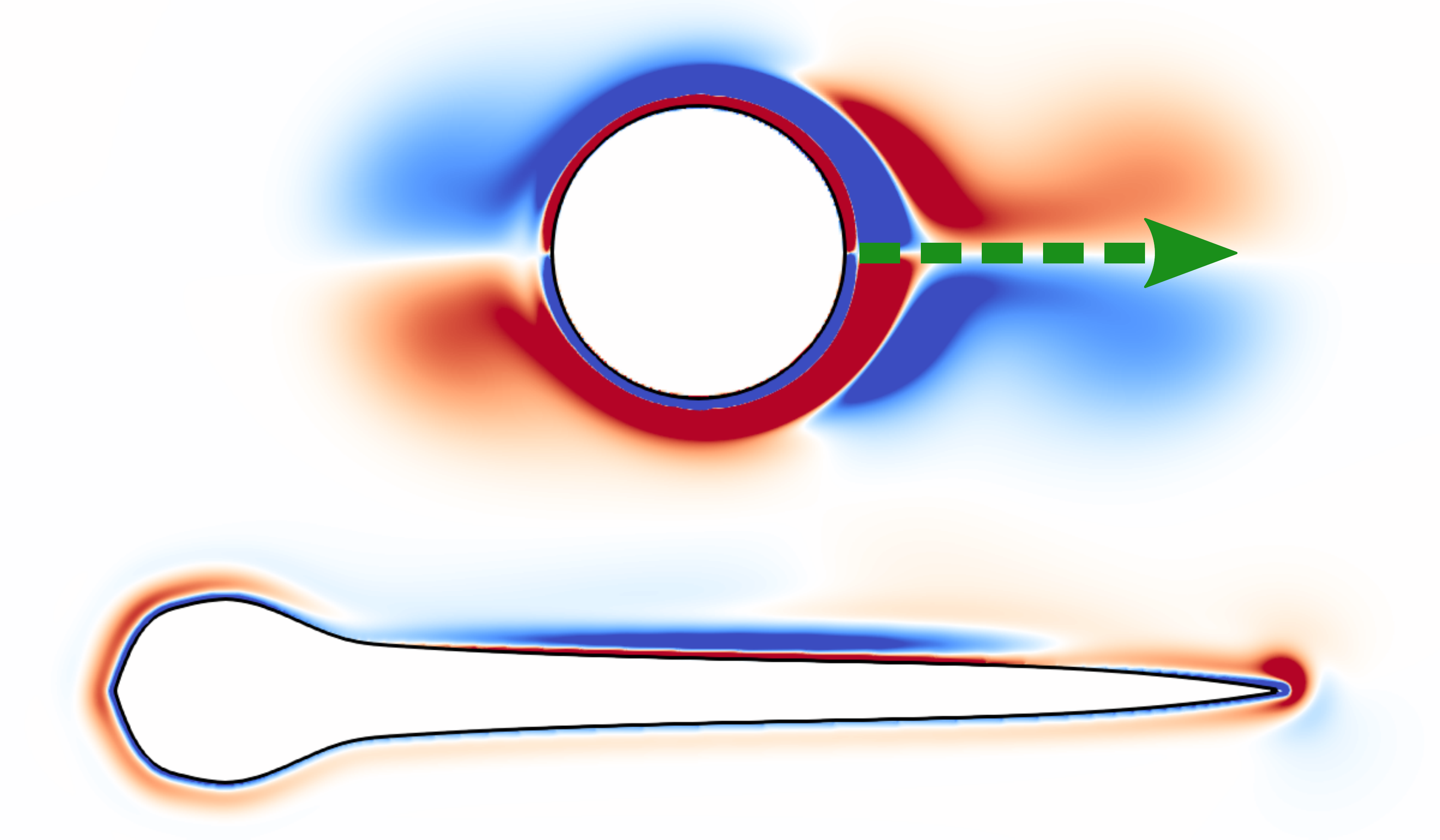}
		\subcaption{}
		\label{fig:horizSnap1}
	   \end{subfigure}
        \begin{subfigure}[b]{0.32\textwidth}
                \centering
		\includegraphics[width=\textwidth]{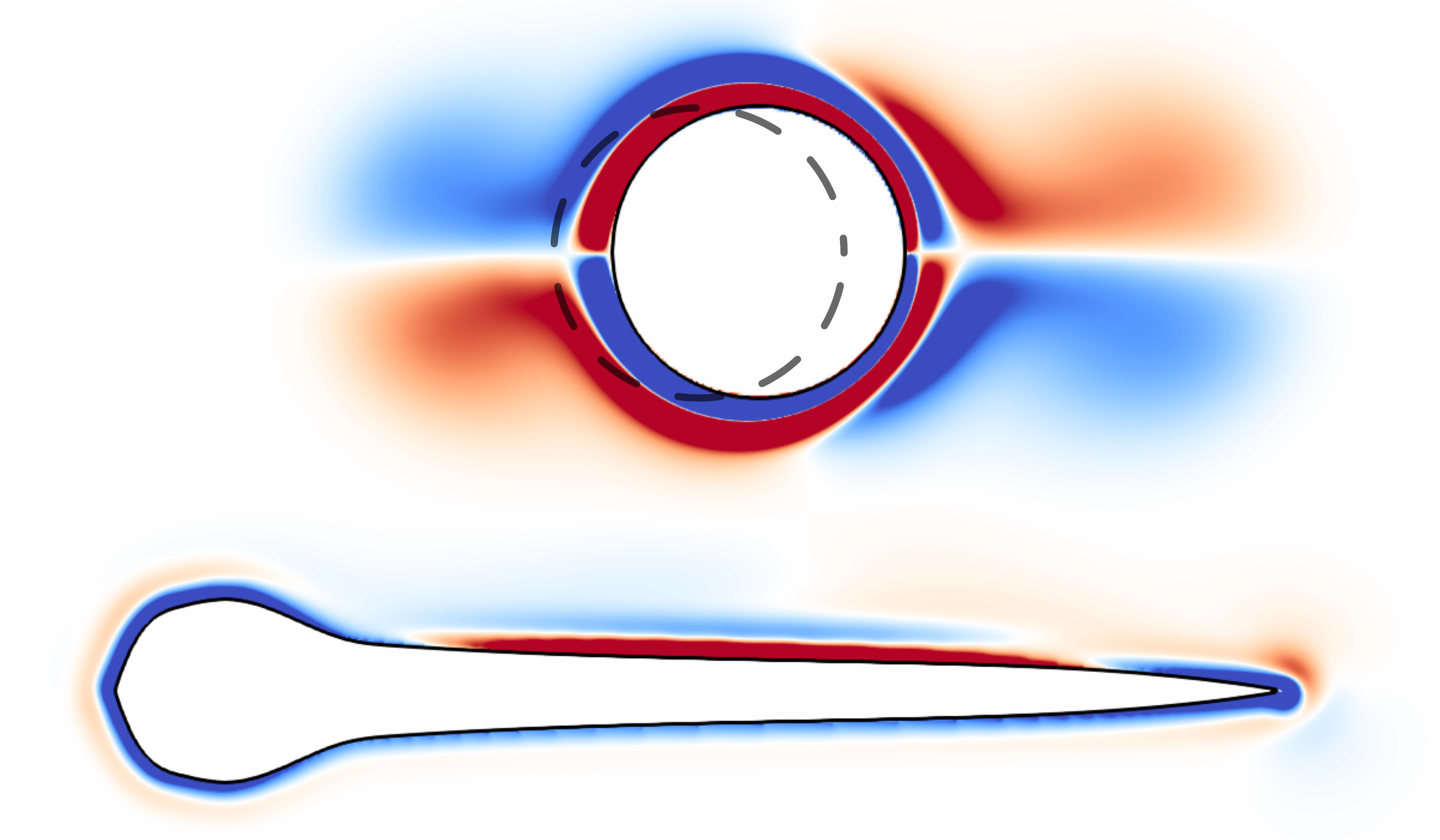}
                \subcaption{}
		\label{fig:horizSnap2}
        \end{subfigure}
        \begin{subfigure}[b]{0.32\textwidth}
                \centering
		\includegraphics[width=\textwidth]{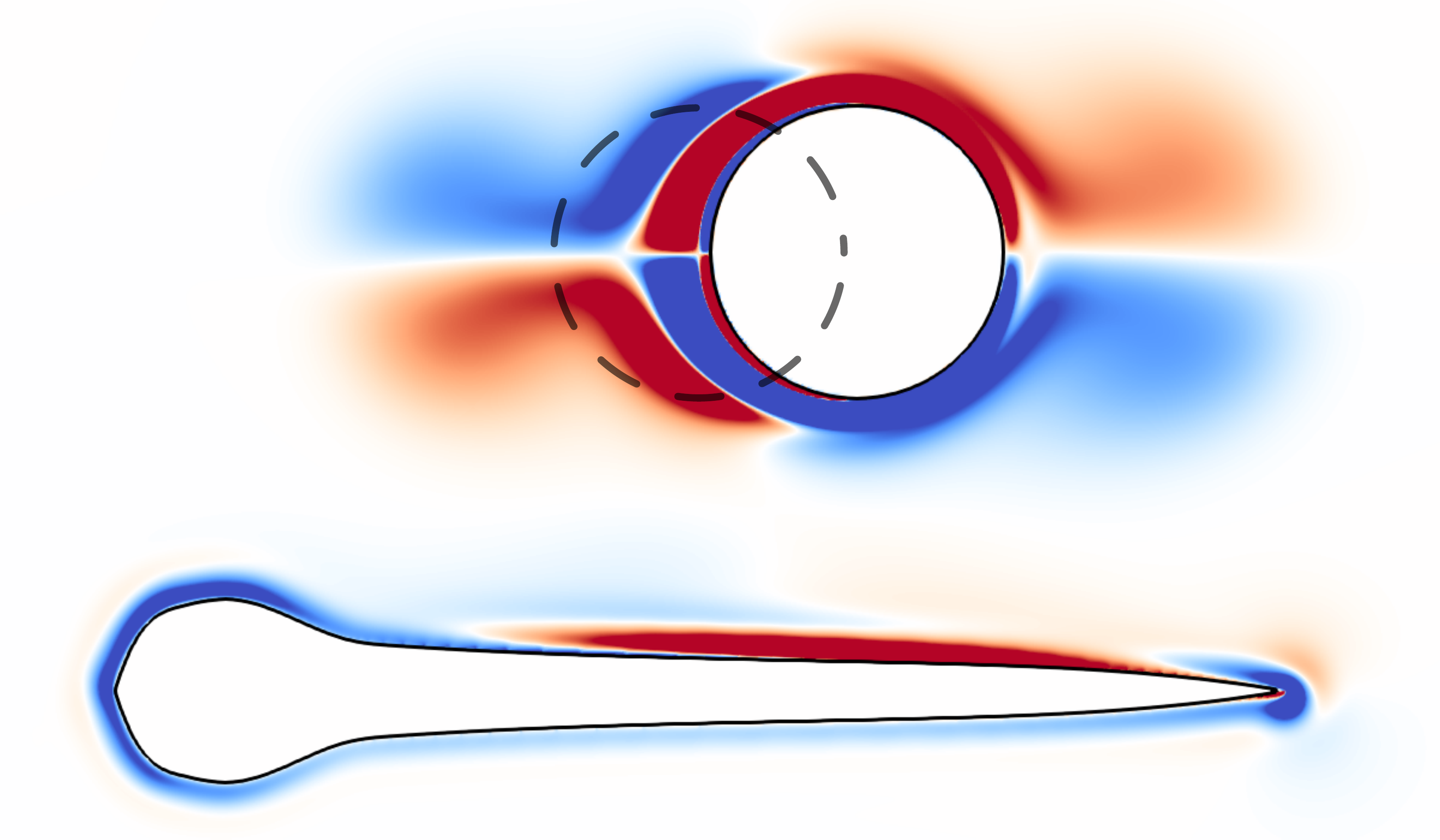}
                \subcaption{}
		\label{fig:horizSnap3}
        \end{subfigure}
        \begin{subfigure}[b]{0.32\textwidth}
                \centering
                \includegraphics[width=\textwidth]{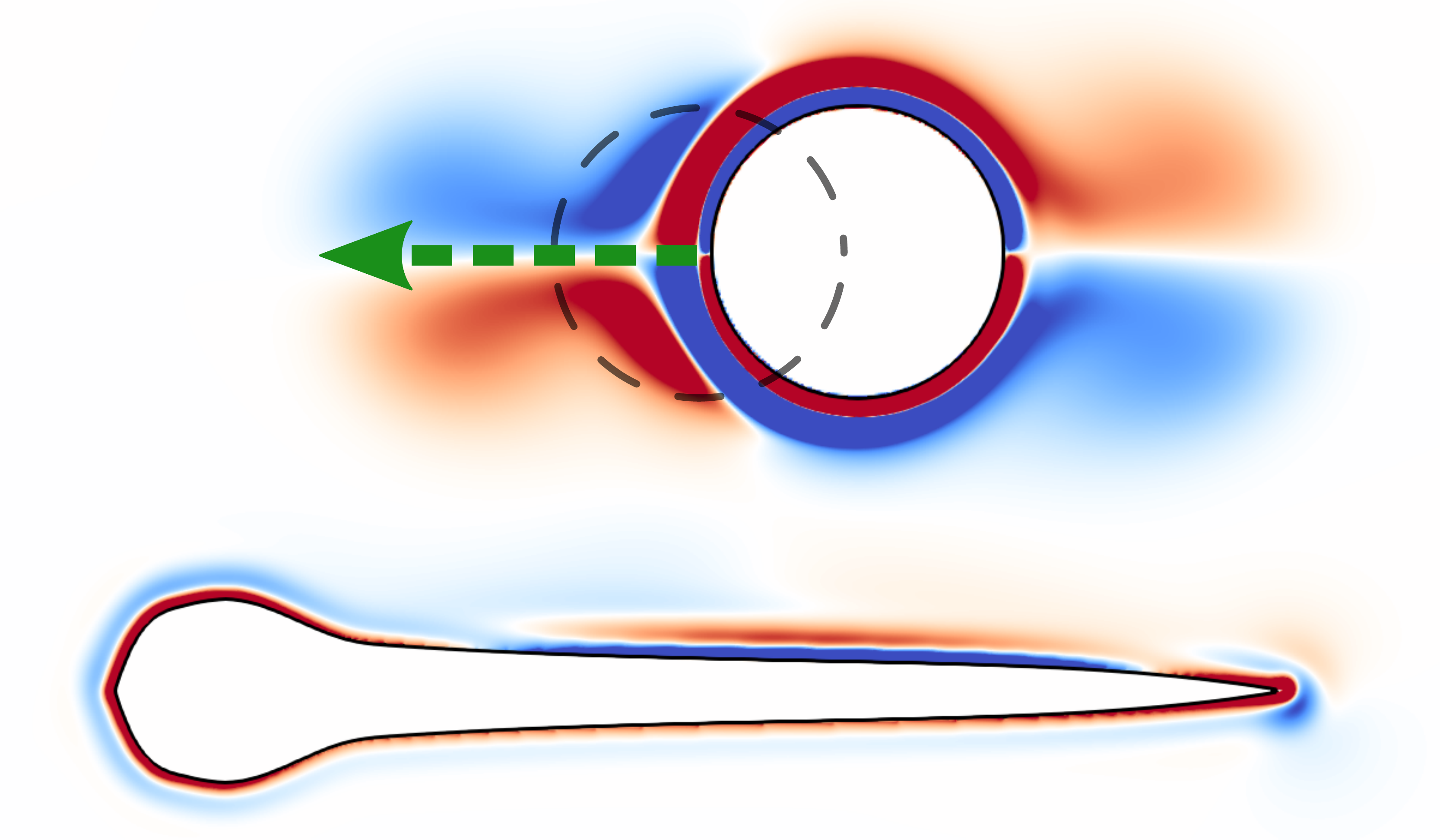}
                \subcaption{}
 		\label{fig:horizSnap4}
        \end{subfigure}
		\begin{subfigure}[b]{0.32\textwidth}
                \centering
                \includegraphics[width=\textwidth]{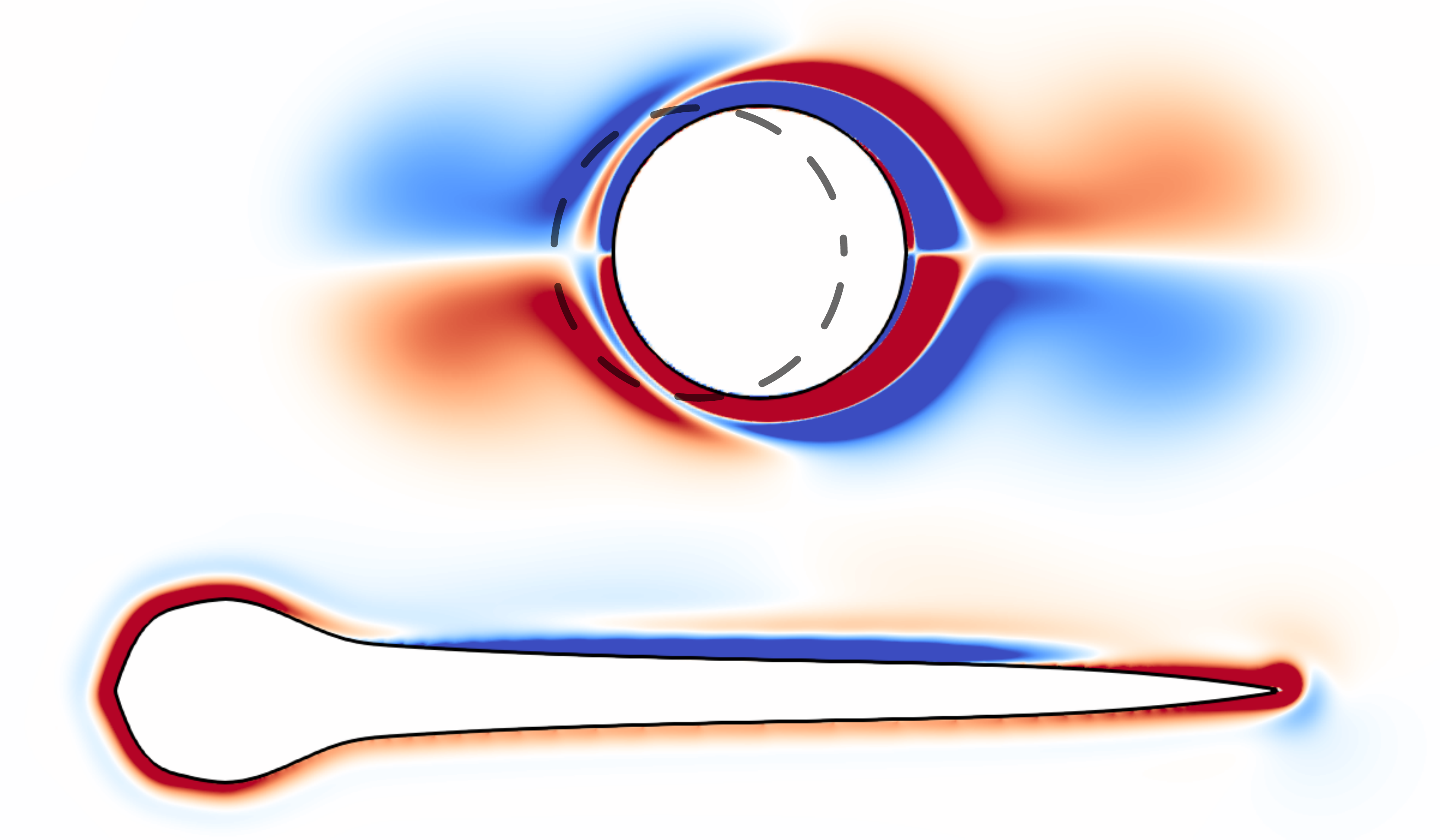}
                \subcaption{}
 		\label{fig:horizSnap5}
        \end{subfigure}
		\begin{subfigure}[b]{0.32\textwidth}
                \centering
                \includegraphics[width=\textwidth]{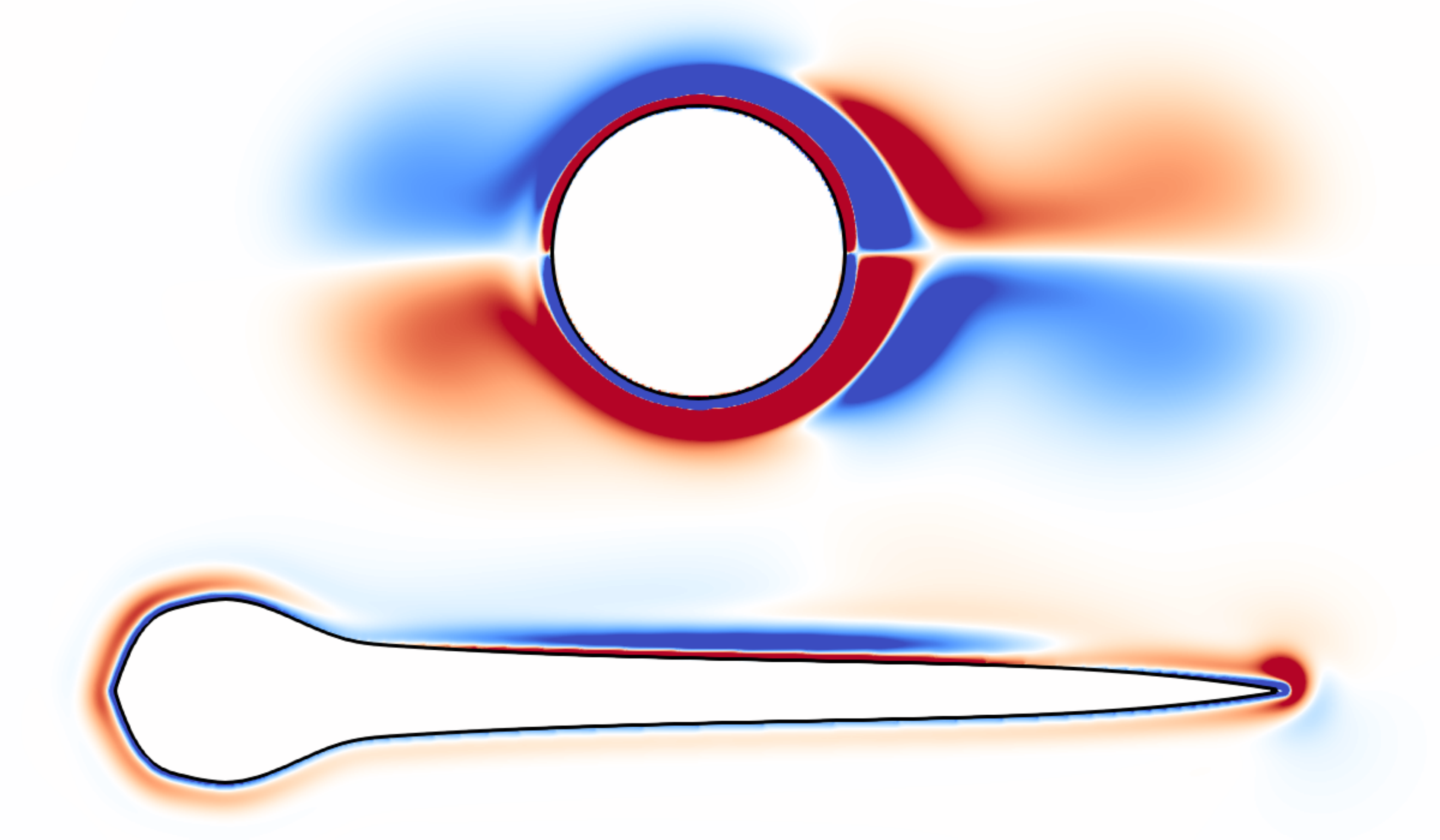}
                \subcaption{}
 		\label{fig:horizSnap6}
        \end{subfigure}
		\caption{Snapshots of the vorticity field around a static larva profile in the presence of a horizontally oscillating cylinder. The snapshots are taken at regular intervals over a single oscillation period, with positive vorticity shown in red and negative vorticity shown in blue. A corresponding animation is shown in \blue{supplementary} \movieHorizCylStatic{}.}
\label{fig:snapshotsOscillating}
\end{figure}

We first consider the  setup  of a stationary  larva-shaped swimmer and a cylinder that either oscillates parallel to the `anteroposterior' axis of the body, or rotates with a constant angular velocity. The oscillating-cylinder setup is shown in Figure~\ref{fig:snapshotsOscillating}, and depicts the vorticity generated  by the cylinder along the larva's body.
These two setups allow us to analyze mechanical cues (i.e., vibrations in the flow field) without interference from a self-generated boundary layer, which has a tendency to obscure external signals in the case of towed and self-propelled bodies. The \blue{simulation domain extends from $[0,1]$ in $x$ and $y$, and the} rectangular prior-region in both the setups corresponds to $\cylinderPos_{min}=(0.357, 0.375)$ and $\cylinderPos_{max}=(0.7, 0.47)$. A total of $11\times 37 = 407$ potential $\cylinderPos^{(i)}$ locations are distributed uniformly throughout the region, and the static object's center of mass \blue{is} located at $(0.5, 0.3)$. \blue{The kinematic viscosity is set to $\nu=1e\text{-}4$ in these  simulations.}

\subsubsection{The utility function, and sensor placement}

\begin{figure}
        \centering
        \begin{subfigure}[b]{0.49\textwidth}
                \centering
                \includegraphics[width=\textwidth]{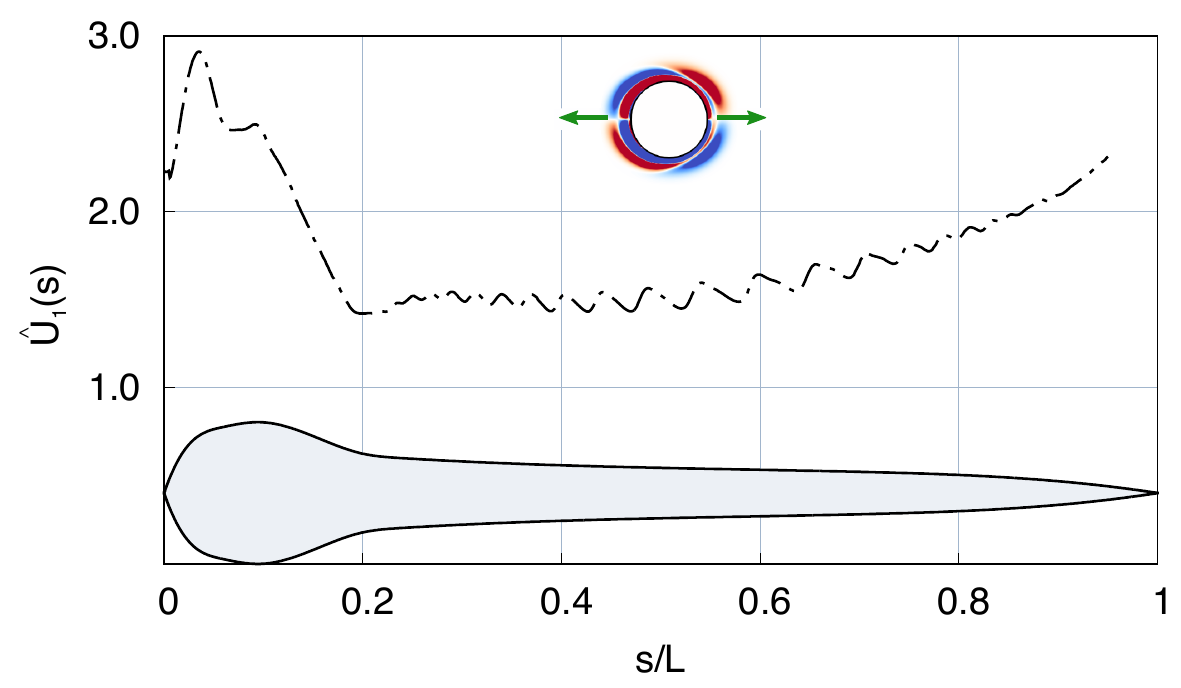}
                \subcaption{}
		\label{fig:utility1dStaticHoriz}
	\end{subfigure}
        \begin{subfigure}[b]{0.49\textwidth}
                \centering
                \includegraphics[width=\textwidth]{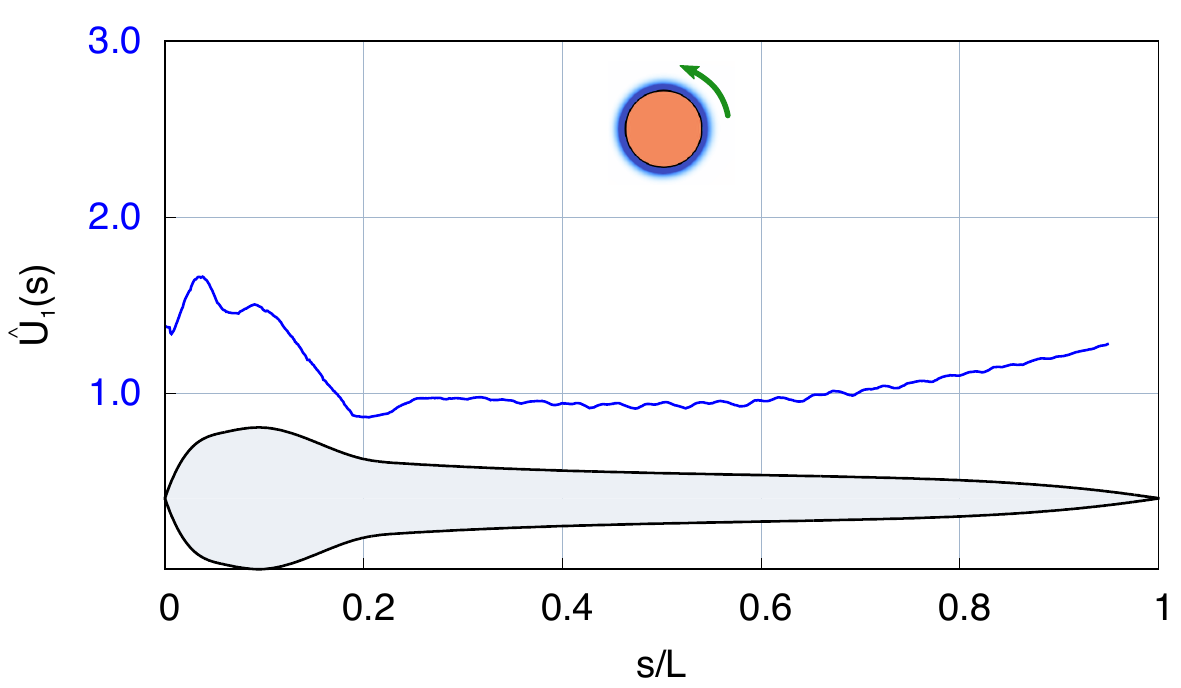}
                \subcaption{}
		\label{fig:utility1dStaticRot}
        \end{subfigure}
        \begin{subfigure}[b]{0.45\textwidth}
                \centering
		\includegraphics[width=\textwidth]{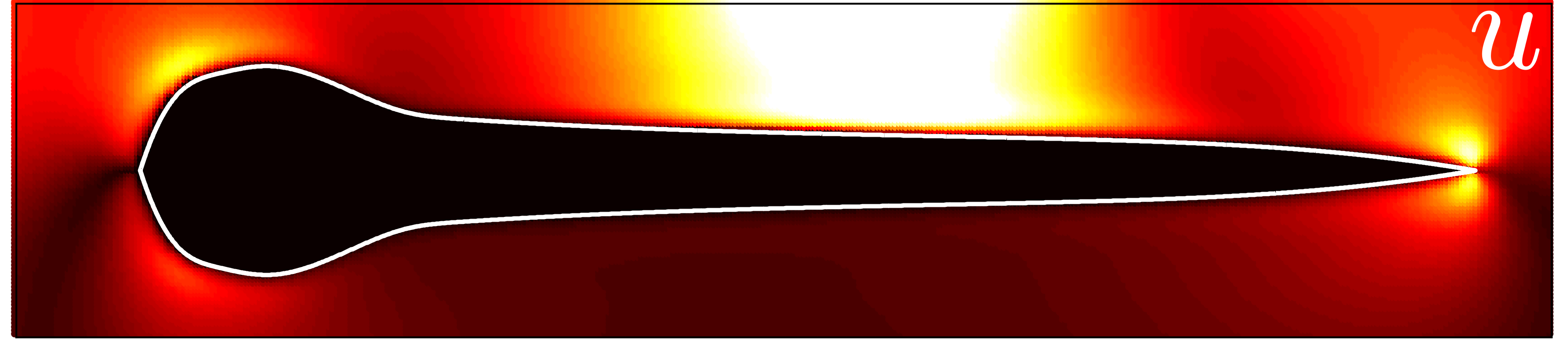}
                \subcaption{}
		\label{fig:stdU_staticHoriz}
        \end{subfigure} \quad
        \begin{subfigure}[b]{0.45\textwidth}
                \centering
                \includegraphics[width=\textwidth]{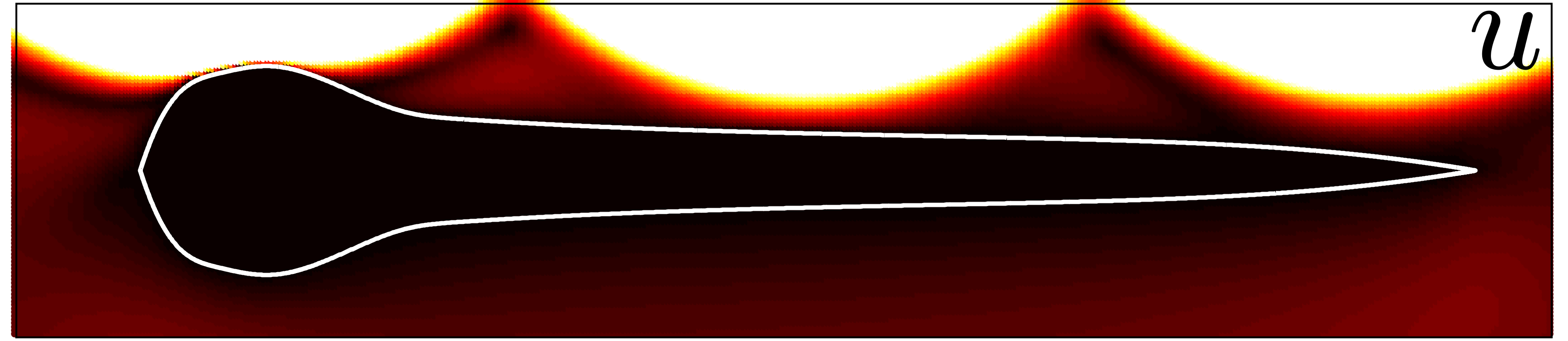}
                \subcaption{}
 		\label{fig:stdU_staticRot}
        \end{subfigure}
                \begin{subfigure}[b]{0.45\textwidth}
                \centering
                \includegraphics[width=\textwidth]{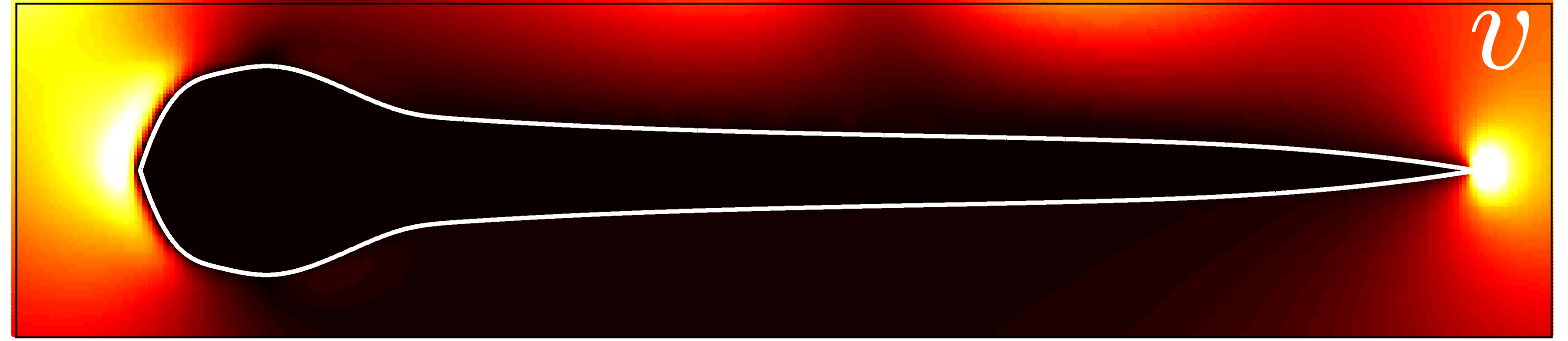}
                \subcaption{}
 		\label{fig:stdV_staticHoriz}
        \end{subfigure} \quad
        \begin{subfigure}[b]{0.45\textwidth}
                \centering
		\includegraphics[width=\textwidth]{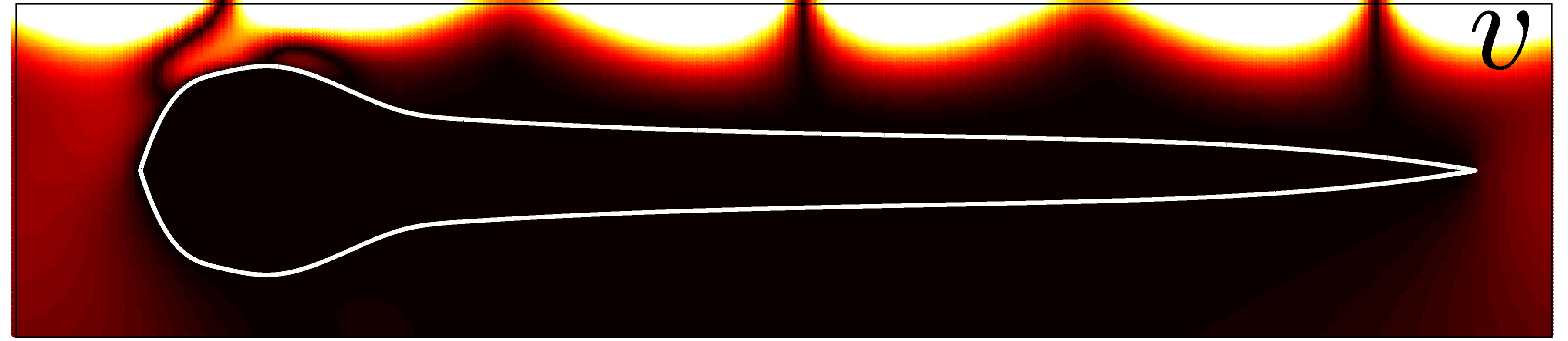}
                \subcaption{}
 		\label{fig:stdV_staticRot}
        \end{subfigure}
         \begin{subfigure}[b]{\textwidth}
                \centering
		\includegraphics[]{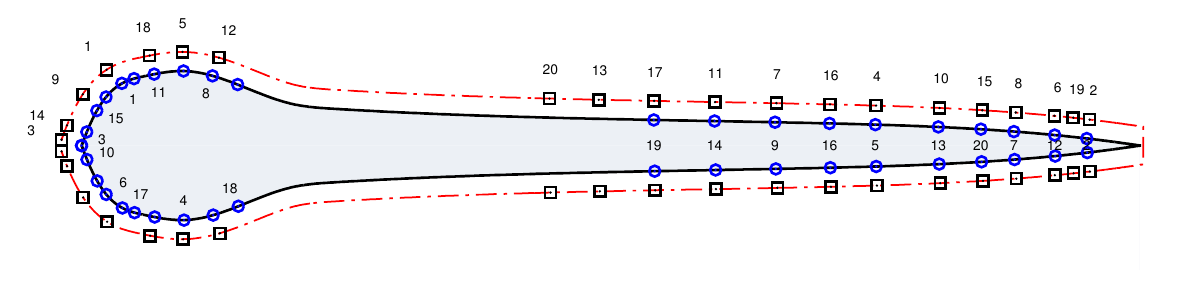}
                \subcaption{}
                \label{fig:staticSensors}
        \end{subfigure} \\
		\caption{Utility plots for a stationary, larva-shaped body with (\subref{fig:utility1dStaticHoriz}) oscillating and (\subref{fig:utility1dStaticRot}) rotating cylinders. The curves indicate the utility for placing the first \blue{shear stress} sensor at a given location $s$. The utility curves were not computed in the region $0.95<s/L\le1$, to avoid potential numerical issues resulting from sharp corners at the tail. (\subref{fig:stdU_staticHoriz},\subref{fig:stdV_staticHoriz}) Standard deviation of horizontal and vertical velocity caused by oscillating cylinders, with larger deviation shown in yellow and lower values shown in black. The standard deviation was computed across 9 distinct simulations (6 time-snapshots recorded in each simulation), with a single oscillating cylinder placed at 9 locations uniformly in the prior-region. (\subref{fig:stdU_staticRot}, \subref{fig:stdV_staticRot}) Standard deviation of velocity components for the rotating cylinders. (\subref{fig:staticSensors}) Optimal sensor distribution determined using sequential placement. Sensors for detecting oscillating cylinders are shown as black squares, whereas those for detecting rotating cylinders are shown as blue circles. The numbering indicates the \blue{sequence determined by the optimal placement algorithm.}}
\label{fig:staticLarva}
\end{figure}

The optimal distribution of sensors along the larva's body can be determined using the estimator $\hat U(\sensors)$ defined in Eq.~\ref{eq:estimator}. Higher utility values indicate that measurements taken at the corresponding locations are more informative. More specifically, the utility at location $s$ is high if a sensor placed there can more effectively differentiate between signals originating from distinct cylinder \blue{locations}. The utility curves computed from signals generated by the oscillating and rotating cylinders are shown in Figures~\ref{fig:utility1dStaticHoriz} and~\ref{fig:utility1dStaticRot}.
The maxima in these curves suggest that the best location for detecting both oscillating and rotating cylinders\blue{, using shear stress sensors,} is at $s/L=0.033$. We remark that this location involves a notable change in body-surface curvature, as can be discerned from the swimmer-silhouettes shown in the figures.

We postulate that the best sensor positions are those that are exposed to large variations in the quantity of interest, namely the shear stress \blue{ or pressure gradient}, since this would allow the sensors to best distinguish between different disturbance sources more readily. We confirm that this is indeed the case, by visualizing the standard deviation of  velocity components in regions surrounding the larva, in Figures~\ref{fig:stdU_staticHoriz} to~\ref{fig:stdV_staticRot}. The standard deviation measures the variation among simulations when cylinders are placed at different positions in the prior-region shown in Figure~\ref{fig:priorRegion}. \refOne{The colour scales are identical for panels~\ref{fig:stdU_staticHoriz} and~\ref{fig:stdV_staticHoriz} (the oscillating cylinder scenario), but different from the colour scales in panels~\ref{fig:stdU_staticRot} and~\ref{fig:stdV_staticRot} (the rotating cylinder scenario). The colour scales in~\ref{fig:stdU_staticRot} and~\ref{fig:stdV_staticRot} have been saturated by approximately 30 times, so that weaker flow disturbances created by the rotating cylinders are adequately visible.}

We observe from Figures~\ref{fig:stdU_staticHoriz} and~\ref{fig:stdV_staticHoriz} that changing the position of an oscillating cylinder gives rise to significant differences in the tangential velocity (shear stress) close to the head and the tail. This implies that signals measured by sensors in these regions differ markedly from one simulation to the other, which \blue{arguably would} make it easier to estimate the position of \blue{a particular}  cylinder. A large variation in horizontal velocity $u$ occurs close to a change in body curvature at $s/L\approx0.033$, which also corresponds to the global maximum in $\hat{U}_1(s)$ (Figure~\ref{fig:utility1dStaticHoriz}). The utility curve exhibits consistently high values for $s/L\le0.15$, which results from large variations in $u$ and $v$ in regions surrounding the head. We note that large variations in the lateral velocity $v$ occur primarily at the head- and tail-tip \blue{(Figure~\ref{fig:stdV_staticHoriz})}, with almost no variation along the midsection ($0.2<s/L\le1$). This can be attributed to $v$ being almost zero in these regions (across all simulations), owing to negligible recirculation  along these relatively straight body sections. The large variation in $v$ at the head/tail tip may be explained by the flow turning at the corners, as is evident from the time-series snapshots shown in Figure~\ref{fig:snapshotsOscillating}. We note that while $u$ appears to exhibit large deviation around the midsection ($0.4\le s/L \le 0.6$ in Figure~\ref{fig:stdU_staticHoriz}), the utility curve in Figure~\ref{fig:utility1dStaticHoriz} does not show a corresponding spike. This may be related to the fact that the standard deviation plots were compiled using a small subset of 9 cylinder-locations out of the 407 used for the utility plot. Furthermore, a close inspection of Figure~\ref{fig:stdU_staticHoriz} indicates that these large deviations in $u$ near the midsection occur beyond the detection range of the sensors, i.e., too far away to be picked up by microscopic neuromasts that are $0.0024L$ in length.

As in the case of oscillating cylinders, the standard deviation plots for rotating cylinders in Figures~\ref{fig:stdU_staticRot} and \ref{fig:stdV_staticRot} can be correlated to the utility curve in Figure~\ref{fig:utility1dStaticRot}; high utility values ($s/L \le 0.15$, Figure~\ref{fig:utility1dStaticRot}) correspond to large deviations in both $u$ and $v$ near the head (Figures~\ref{fig:stdU_staticRot} and~\ref{fig:stdV_staticRot}). Based on the utility curve, the highest sensitivity for measuring flow perturbations corresponds to the head and posterior sections of the body. This suggests that the head and tail are the most informative regions for \blue{detecting shear stress fluctuations for a static larva}, regardless of the type of disturbance being considered. This observation is consistent  with the distribution of neuromasts shown in Figure~\ref{fig:fishNeuromasts}, where the \blue{surface neuromasts} are visible in high concentrations in the head and posterior regions of fish, but show sparse presence along the midsection. 

\subsubsection{Sequential sensor placement}

In the previous section we discussed the case of a single sensor on the swimmer body. We now examine the optimal arrangement of multiple sensors, where the best location for the $n$-th sensor is determined provided that $n-1$ sensors have already been placed. Assume that the first sensor has been placed at $s_1^\star$ using the global maximum in utility curve $\hat{U}_1(s)$. The next best sensor-location is determined by recomputing the utility function $\hat{U}_2(s)$ as described in section~\ref{sec:sequentialOpt}. Following this procedure, the optimal location of all  sensors is determined sequentially.

Figure~\ref{fig:staticSensors} shows the optimal distribution of 20 sensors for the static larva determined in this manner. We first examine the optimal arrangement for detecting oscillating cylinders, with the corresponding sensors depicted as black squares. We observe that out of the first 10 sensors, numbers $\{1, 3, 5, 9\}$ are placed at the head, whereas numbers $\{2, 4, 6, 7, 8, 10\}$ are found towards the posterior. This suggests a large  information-gain via sensors located in the head \blue{and the tail.}  For detecting rotating cylinders (sensors shown as blue circles in Figure~\ref{fig:staticSensors}), sensors $\{1, 3, 4, 6, 8, 10\}$ are found in the head, and sensors $\{2, 5, 7, 9\}$ are placed in the posterior section.
%
%

\begin{figure}
        \centering
        \begin{subfigure}[c]{0.48\textwidth}
                \centering
		\includegraphics[width=\textwidth]{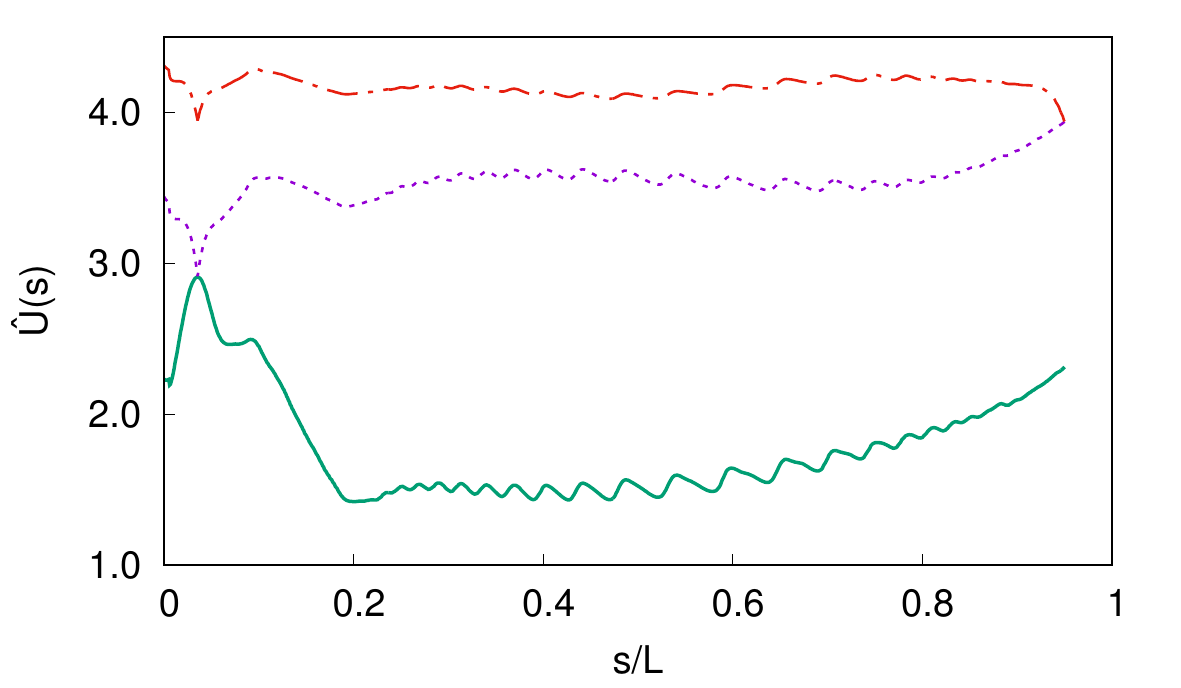}
                \subcaption{}
                \label{fig:utility_successive}
        \end{subfigure} 
        \begin{subfigure}[c]{0.48\textwidth}
                \centering
                \includegraphics[width=\textwidth]{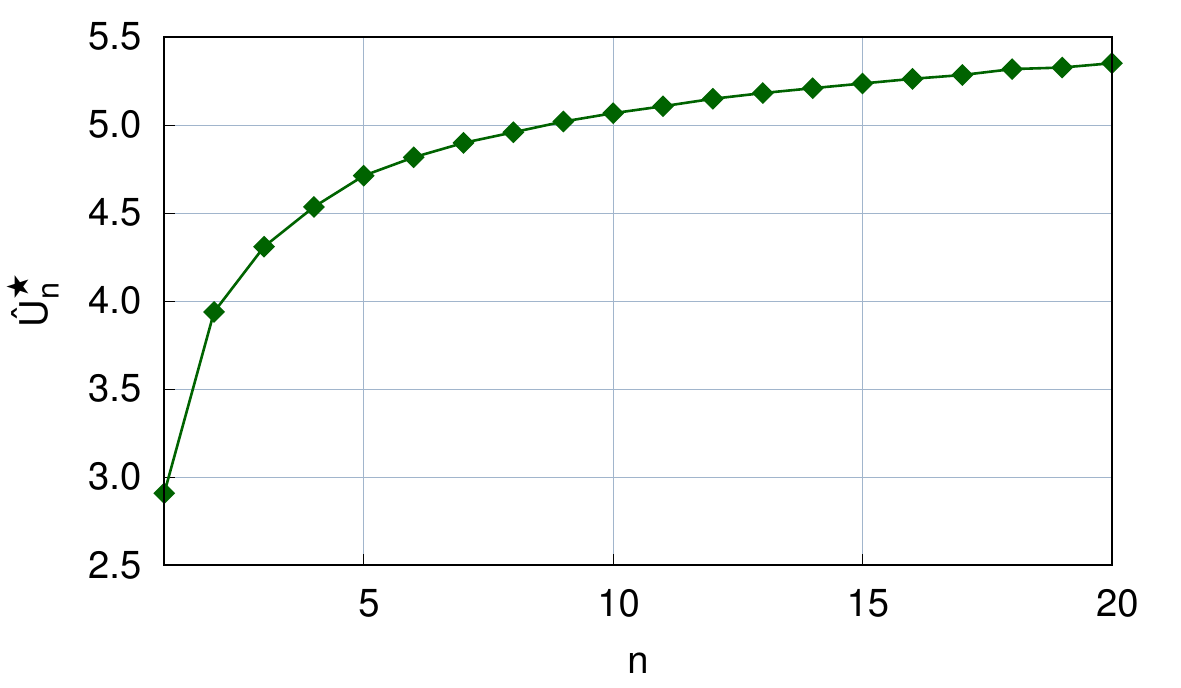}
                \subcaption{}
                \label{fig:utility_cumulative}
        \end{subfigure}
	\caption{(\subref{fig:utility_successive}) Utility curves for placing the first three sensors on a static larva that detects oscillating cylinders (Figure~\ref{fig:snapshotsOscillating}). The solid green curve corresponds to $\hat{U}_1(s)$, the dashed purple curve to $\hat{U}_2(s)$, and the red dash-dot curve to $\hat{U}_3(s)$. (\subref{fig:utility_cumulative}) The optimal utility $\hat U_{n}^\star$ for the $n$-th sensor can be determined as $\max_{s} \hat{U}_n(s)$ using the curves shown in panel (\subref{fig:utility_successive}) (see also Eq.~\ref{eq:optimal:sensor}).}
\label{fig:utility_sequential}
\end{figure}

We also examine the utility curves for placing the first three \blue{oscillation-detecting shear stress} sensors in Figure~\ref{fig:utility_successive}.
We observe that $\hat{U}_2(s_1) \approx \hat{U}_1(s_1)$, which indicates that placing a second sensor at the same location as the first ($s_1/L=0.033$) would not lead to an appreciable increase in the utility value (i.e., no gain in useful information). The maximum in $\hat{U}_2(s)$ occurs at $s/L=0.95$, which yields the optimal location $s_2^\star$ for the second sensor. Another notable aspect of curve $\hat{U}_2(s)$ is a pronounced `v-shaped' depression in the vicinity of $s_1^\star$, which results from using a non-zero correlation length in Eq.~\ref{eq:covarianceMatrix}. The low utility values in this region impede the placement of sensors too close \blue{to each other}. Using a zero correlation length would have resulted in an abrupt drop in $\hat{U}_2(s)$ at $s_1^\star$ (instead of the smooth depression), and could lead to excessive clustering of sensors within confined neighbourhoods. Figure~\ref{fig:utility_cumulative} shows the cumulative utility value for an increasing number of sensors placed on the swimmer body. We observe that after a rapid initial rise for the first three to five sensors, the utility of placing subsequent sensors increases very slowly. This indicates that using a limited number of  optimal \blue{sensor} locations should be sufficient to characterize disturbance sources with \blue{reasonably good}  accuracy.

\subsection{Motionless larva in the wake of a D-cylinder}
\begin{figure}
        \centering
        \begin{subfigure}[b]{0.48\textwidth}
                \centering
                \frame{\includegraphics[width=\textwidth]{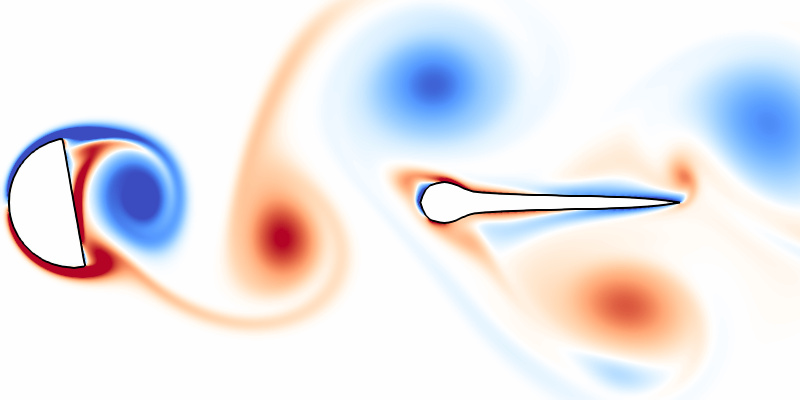}}
                \subcaption{}
		\label{fig:dCylSnap1}
	   \end{subfigure}
        \begin{subfigure}[b]{0.48\textwidth}
                \centering
                \frame{\includegraphics[width=\textwidth]{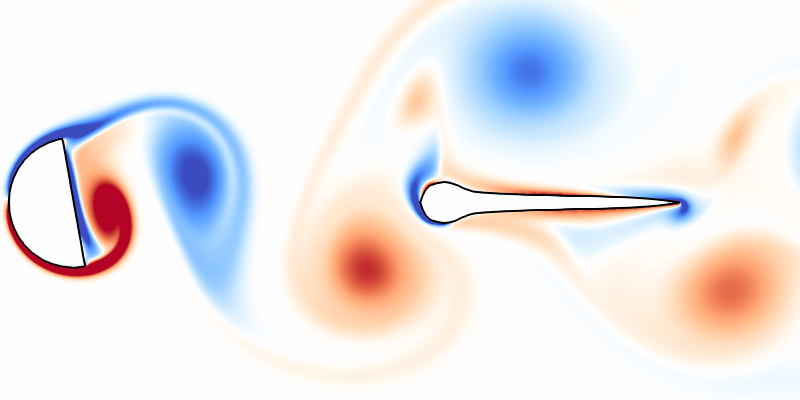}}
                \subcaption{}
		\label{fig:dCylSnap2}
        \end{subfigure}
        \begin{subfigure}[b]{0.48\textwidth}
                \centering
		\frame{\includegraphics[width=\textwidth]{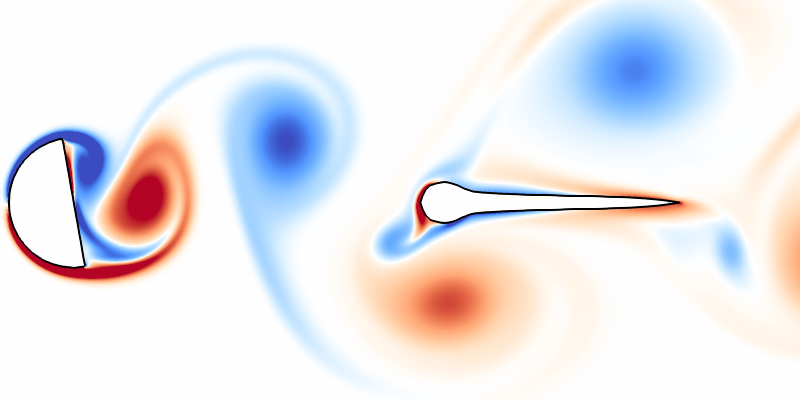}}
                \subcaption{}
		\label{fig:dCylSnap3}
        \end{subfigure}
        \begin{subfigure}[b]{0.48\textwidth}
                \centering
                \frame{\includegraphics[width=\textwidth]{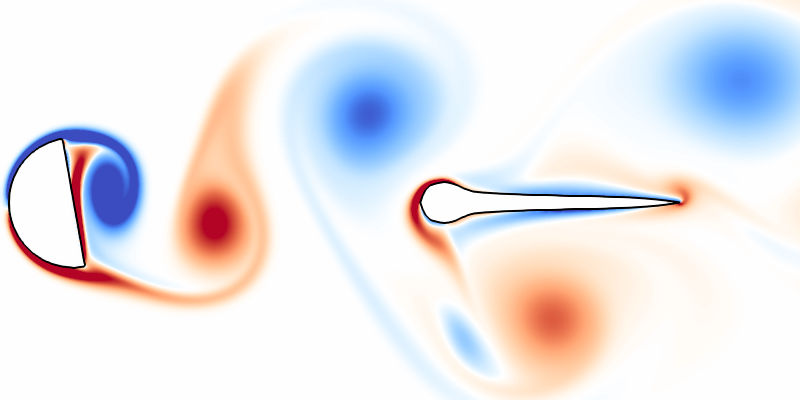}}
                \subcaption{}
 		\label{fig:dCylSnap4}
        \end{subfigure}
		\caption{Snapshots of the vorticity field around a static larva in the wake of a D-shaped cylinder with diameter $0.5L$. The D-cylinder is oriented at a {10\degree} angle with respect to a uniform horizontal flow  to promote vortex shedding. The snapshots are shown at regular time-intervals, with positive vorticity shown in red and negative vorticity shown in blue. A corresponding animation is show in \movieDcyl.}
\label{fig:dCylSnapshots}
\end{figure}

We now consider simulations where a rigid larva-shaped profile is placed in the unsteady vortex-wake generated by a D-shaped half cylinder (Figure~\ref{fig:dCylSnapshots}).
This configuration is inspired by the pioneering work of Liao et al. ~\citep{Liao2003Science} who examined the fluid dynamics of trout placing themselves  behind rocks. A uniform horizontal flow of $1L/s$ is imposed throughout the computational domain, and the rigid bodies are held stationary. The D-cylinder is located at $(0.2,0.5)$, and the rectangular prior-region for placing the larvae extends from $\cylinderPos_{min}=(0.3, 0.43)$ to $\cylinderPos_{max}=(0.79, 0.57)$. A total of $11\times 36 = 396$ potential $\cylinderPos^{(i)}$ locations are distributed uniformly throughout the prior-region. The Reynolds number is $Re=200$ based on the cylinder diameter, and $Re=400$ based on the swimmer length. 

\begin{figure}
        \centering
        \begin{subfigure}[c]{\textwidth}
                \centering
				\includegraphics[width=0.8\textwidth]{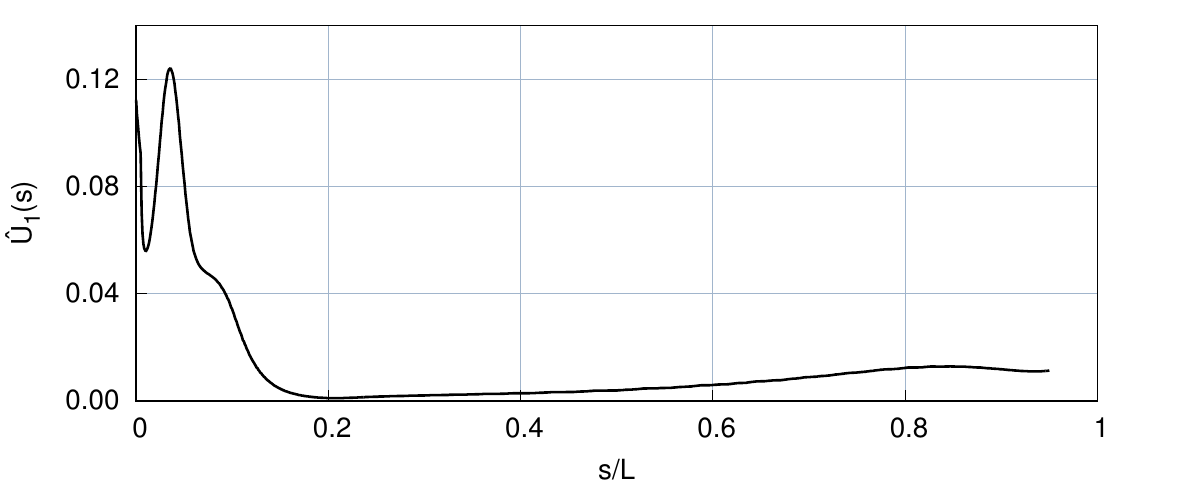}
		\subcaption{}
                \label{fig:utilityDcyl}
        \end{subfigure} \\
	\begin{subfigure}[c]{\textwidth}
                \centering
                \includegraphics[width=0.8\textwidth]{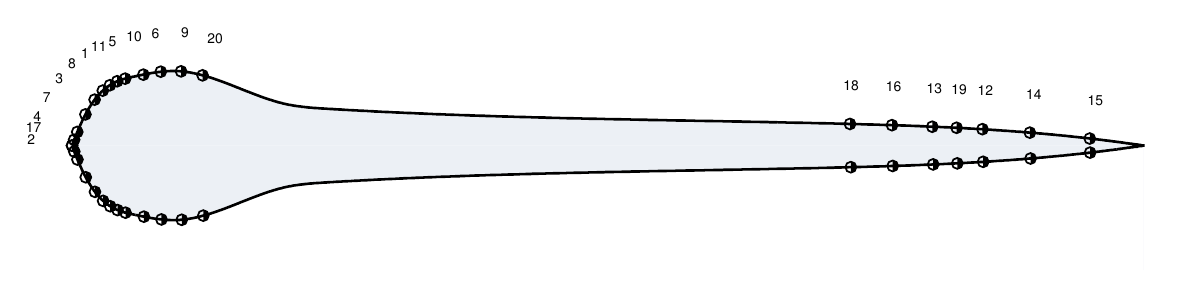}
                \subcaption{}
                \label{fig:sensorsDcyl}
        \end{subfigure} 
                \caption{(\subref{fig:utilityDcyl}) Utility curve $\hat{U}_1(s)$ for larvae in a D-cylinder's wake. (\subref{fig:sensorsDcyl}) Sequential placement of 20 sensors, with the order of placement shown.}
\label{fig:dCylSensors}
\end{figure}

Figure~\ref{fig:dCylSensors} shows the utility curve for placing the first shear stress sensor on the \blue{static} larva, as well as the sensor distribution resulting from sequential placement.
 The utility values for  $0.2 \le s/L \le 0.6$ are close to zero, which implies that placing the first sensor along the midsection would provide minimal information gain. Using the sequential-placement procedure described in section~\ref{sec:sequentialOpt}, we determine that \blue{all}  of the first 10 sensors are placed at the head, \blue{with no sensors present in the tail.}  Our results indicate that sensors at the head are far more significant  than sensors in the mid- and posterior-sections of the body for detecting the unsteady wake behind a half-cylinder.

\subsection{Self-propelled swimmers\blue{: shear stress sensors}}
\label{sec:swimmingFish}

Fish generate vorticity on their bodies by  their undulatory motion.  Their  flow-sensing neuromasts are completely immersed in this self-generated flow field, which likely has a significant impact on their ability to detect external disturbances. To include the influence of these self-generated flows on optimal sensor-placement, we now consider simulations of self-propelled swimmers that are exposed to oscillating and rotating cylinders (Figure~\ref{fig:showShape}). These swimmers utilize an intermittent swimming gait referred to as `burst-and-coast' swimming, which allows for \blue{improved} sensory perception \citep{Kramer2001}, as self-generated disturbances subside during the coasting phase. The swimmers perform four full burst-coast swimming cycles starting from rest, before the cylinder starts oscillating or rotating, as depicted in \movieRotCyl{} and \movieHorizCyl{}. In the initial transient phase, the swimmer gain a speed of approximately $\blue{0.7}L/s$, which corresponds to a Reynolds number of $\text{Re} =uL/\nu\approx \blue{280}$ (with $L=0.2$ and $\nu=1e\text{-}4$). At the start of the fifth coasting phase, the cylinder starts moving, which simulates the startle/attack response of a prey/predator present in the swimmer's vicinity. The rectangular prior-region for initializing the cylinders extends from $\cylinderPos_{min}=(0.25, 0.375)$ to $\cylinderPos_{max}=(0.7, 0.5)$, with a total of $11\times 37 = 407$ potential $\cylinderPos^{(i)}$ locations distributed uniformly throughout the region. The swimmer's center of mass \blue{is} located at $(0.5, 0.3)$.

\begin{figure}
        \centering
        \begin{subfigure}[c]{\textwidth}
                \centering
				\includegraphics[width=0.8\textwidth]{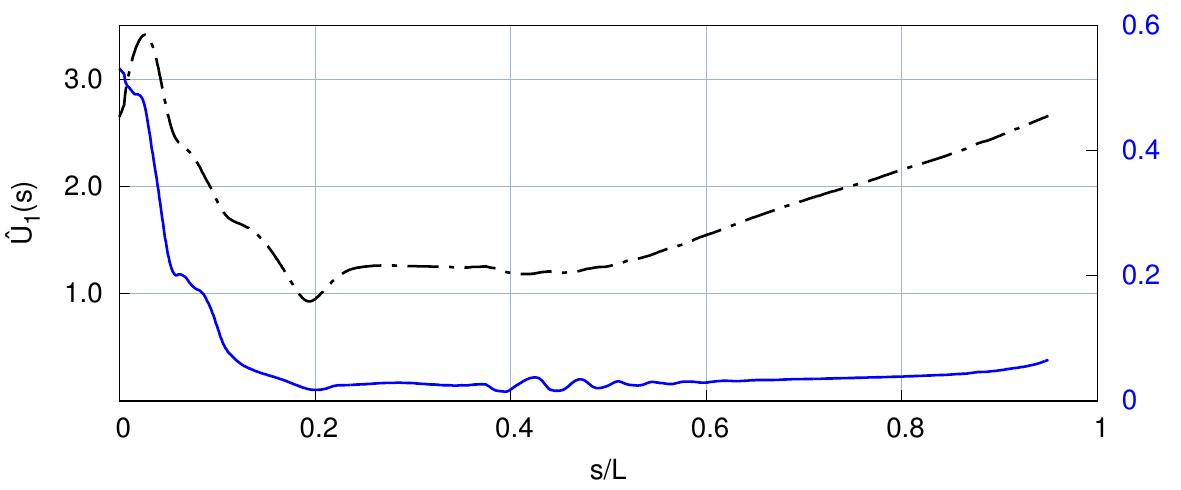}
                \subcaption{}
                \label{fig:utilityLarva}
        \end{subfigure} \\
	\begin{subfigure}[c]{\textwidth}
                \centering
                \includegraphics[width=0.8\textwidth]{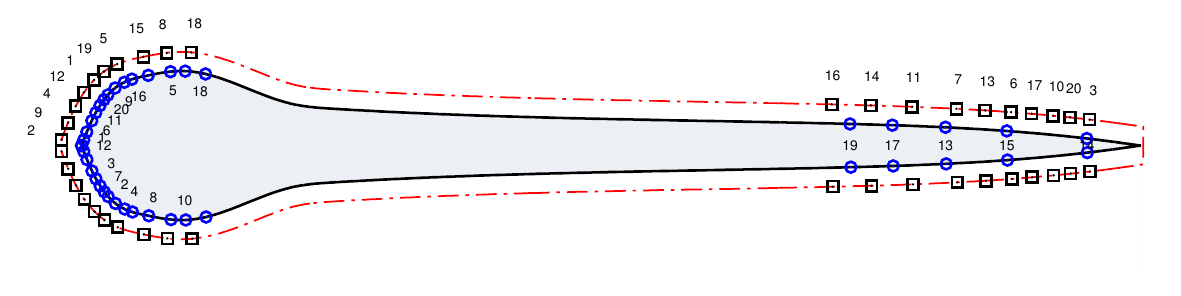}
                \subcaption{}
                \label{fig:sensorsLarva}
        \end{subfigure} \\
	        \begin{subfigure}[c]{\textwidth}
                \centering
                \includegraphics[width=0.8\textwidth]{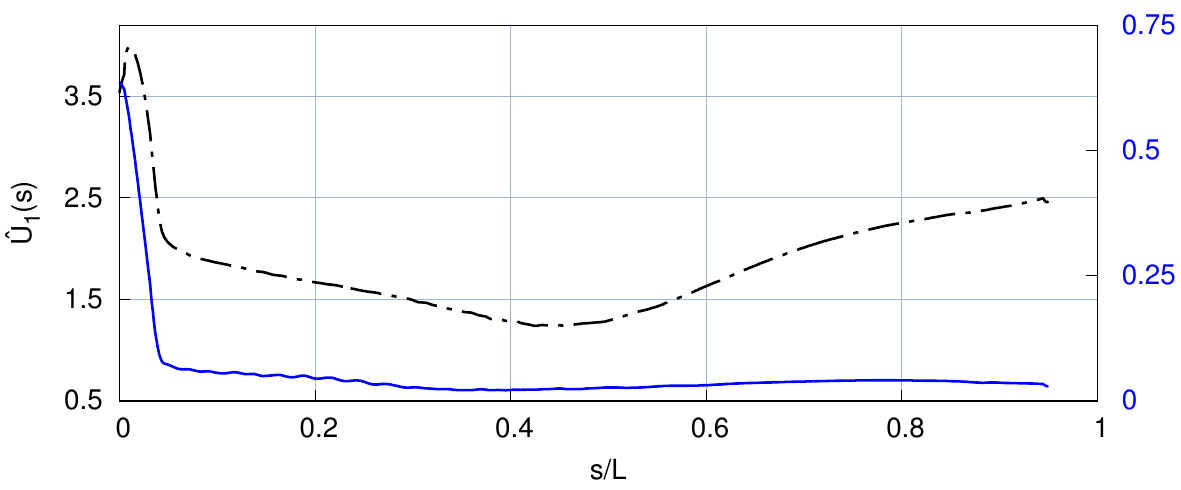}
                \subcaption{}
                \label{fig:utilityAdult}
        \end{subfigure} \\
        \begin{subfigure}[c]{\textwidth}
                \centering
                \includegraphics[width=0.8\textwidth]{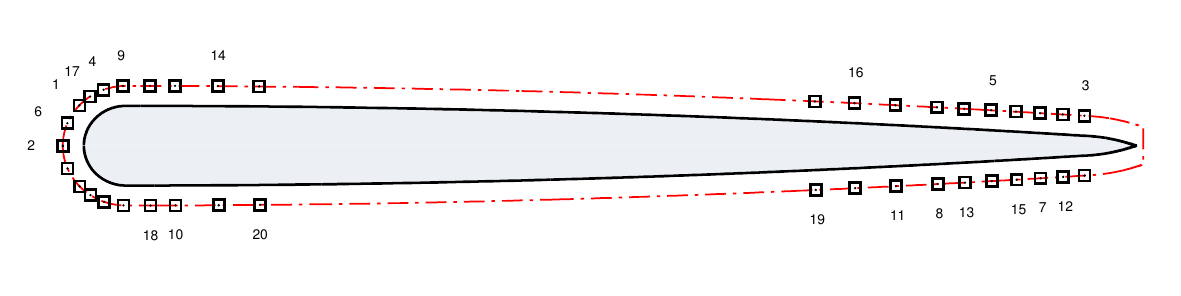}
                \includegraphics[width=0.8\textwidth]{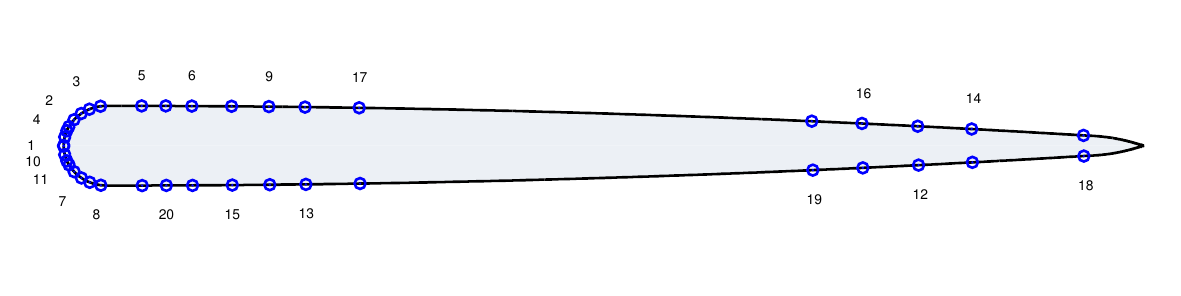}
                \subcaption{}
                \label{fig:sensorsAdult}
        \end{subfigure}
		\caption{(\subref{fig:utilityLarva}) Utility curves for the first \blue{shear stress} sensor, $\hat{U}_1(s)$, on a larva-shaped swimmer (black squares - oscillating cylinders,  blue circles - rotating cylinders). (\subref{fig:sensorsLarva}) Sequential placement of 20 sensors along the body, with the order of placement shown. (\subref{fig:utilityAdult})  Utility curves for an adult-shaped swimmer. (\subref{fig:sensorsAdult}) Sensor placement for the adult, with results from horizontal and rotating disturbances shown separately for clarity.}
\label{fig:utilitySwimmers}
\end{figure}

To determine the extent to which body shape influences optimal placement, we perform simulations using a larva-shaped profile, and a simplified model of an adult. Figure~\ref{fig:utilitySwimmers} compares the utility curves and sensor distributions for these two distinct swimmers.
Based on the utility curves in Figures~\ref{fig:utilityLarva} and~\ref{fig:utilityAdult}, we deduce that the head is the most suitable region for placing the first sensor, as was the case for the motionless profiles examined in the previous sections. \blue{We also observe that the utility curves are correlated to the surface curvature of their respective body profiles; in the case of the larva, there is marked variation in $\hat{U}_1(s)$ for $s/L\le0.2$, which corresponds to large curvature changes in the body surface. The utility curve also shows a gradual variation for $s/L\ge0.6$, which corresponds to a gentler change in curvature of the surface. Similarly, the utility curves and body curvature for the adult vary rapidly for $s/L\le0.05$ and more gradually for $s/L\ge0.6$.} Furthermore, we \blue{note that the blue utility curves in Figures~\ref{fig:utilityLarva} and~\ref{fig:utilityAdult} are close to $0$ for $s/L \ge 0.2$. This suggests that the head is the most useful region for placing rotation-detecting shear stress sensors, irrespective of differences in body shape.} \ref{fig:utilityLarva} ~\ref{fig:utilityAdult}

The optimal sensor arrangements observed in Figures~\ref{fig:sensorsLarva} and~\ref{fig:sensorsAdult} are listed in Table~\ref{tab:swimmerSensors}.
\begin{table}
	\begin{center}
		\def~{\hphantom{0}}
		\begin{tabular}{lllcl}
			&     &Head &Midsection &Posterior \\
			\hline
			\multirow{2}{*}[-1.5pt]{Larva}  &Oscillating	&   $1, 2, 4, 5, 8, 9$	&\textemdash	& $3, 6, 7, 10$ \\[5pt]
				  \cline{2-5}\\[-5pt]
							&Rotating	&$1, 2, \ldots, 10	 $  &\textemdash   		&\textemdash\\
			\hline
			\multirow{2}{*}{Adult}  	&Oscillating  	& $1, 2, 4, 6, 9, 10$ 	&\textemdash	& $3, 5, 7, 8$ \\[5pt]
				  \cline{2-5} \\[-5pt]
							&Rotating 	&$1, 2, \ldots, 10$ 	&\textemdash	&\textemdash\\
			\hline
		\end{tabular}
		\caption{Optimal distribution of the first 10 \blue{shear stress} sensors for the self-propelled swimmers. The body has been divided into 3 distinct segments: the head ($0\le s/L <0.2$); the midsection ($0.2\le s/L < 0.6$); and the posterior ($0.6\le s/L \le1$).}
		\label{tab:swimmerSensors}
	\end{center}
\end{table}
\blue{There is strong indication that the head is the most important region for detecting shear stress caused by external disturbances, followed by the posterior section; the midsection appears to be insensitive to shear-stress variations altogether, as evidenced by the lack of sensors in this region. Moreover, the posterior section appears to be insensitive to rotating disturbances regardless of the body shape.}
\blue{These observations}  agree well with \blue{surface} neuromast distributions observed in live fish (Figure~\ref{fig:lateralLineSystem}), where large numbers are found in the head \blue{and the tail}, with sparser clustering in the  midsection.

\subsection{Optimal sensor placement using combined datasets}
\label{sec:combined_shear}

\begin{figure}
        \centering
        \begin{subfigure}[c]{\textwidth}
                \centering
		\includegraphics[width=0.8\textwidth]{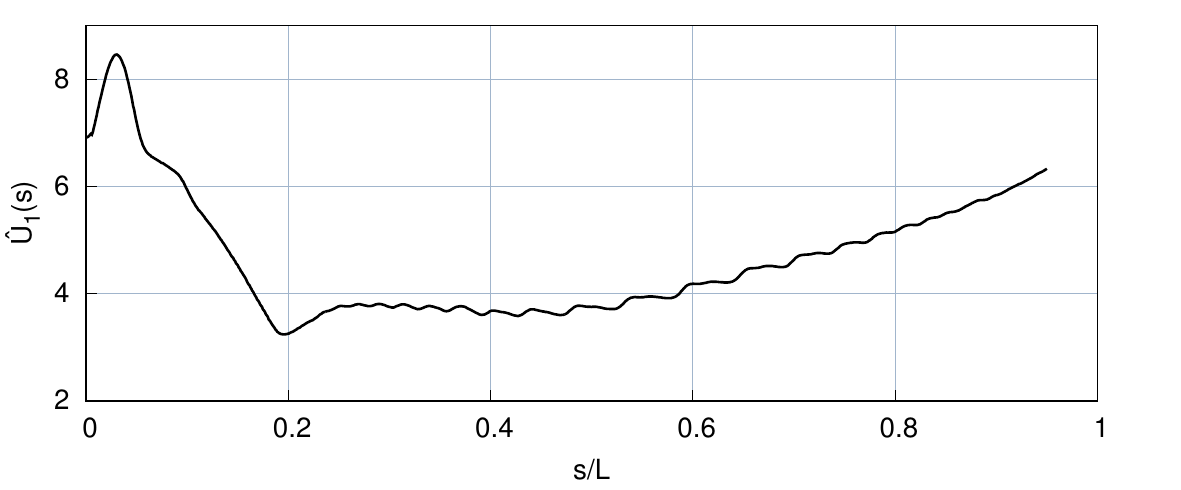}
                \subcaption{}
                \label{fig:utilityAllSum}
        \end{subfigure} \\
	\begin{subfigure}[c]{\textwidth}
                \centering
                \includegraphics[width=0.8\textwidth]{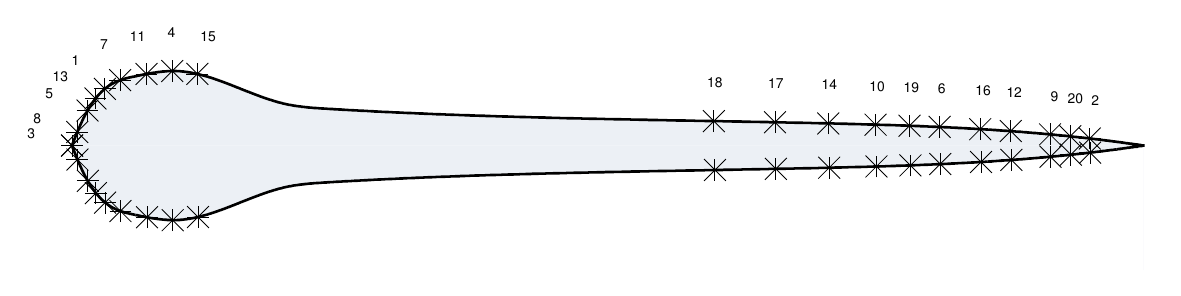}
                \subcaption{}
                \label{fig:sensorsAllSum}
        \end{subfigure} 
		\caption{(\subref{fig:utilityAllSum}) Utility curve for the first \blue{shear stress} sensor, $\hat{U}_1(s)$,  on a larva-shaped swimmer, using a combination of all five flow configurations described in the paper. (\subref{fig:sensorsAllSum}) Sequential placement of 20 \blue{shear stress} sensors along the body.}
\label{fig:combination}
\end{figure}

Fish are subject to a multitude of  external stimuli over the course of their lifetime. Hence, it is conceivable that  neuromasts may be  attuned to diverse  sources of disturbance. We emulate this situation for optimal sensor placement by considering data collected from the five different simulation setups \blue{simultaneously}, namely, motionless larvae with oscillating, rotating, and D-shaped cylinders, and self-propelled larvae with oscillating and rotating cylinders. The sequential-placement procedure described in the previous sections is followed, with a slight modification to the definition of the utility function. The combined utility for the five different configurations may be expressed as a sum of the individual utility functions\blue{,} due to the conditional independence of the measurements on the sensor locations (see Appendix~\ref{sec:app:derivation}).
The resulting utility curve \blue{for the shear stress sensors} is shown in Figure~\ref{fig:combination}, along with the optimal sensor distribution.
We observe predominant placement of sensors in the head \blue{and tail}, corresponding to large utility values \blue{in these regions. Moreover, we find that virtually no sensors are located in the midsection.}  The dense clustering of sensors in the head \blue{and tail}, with sparse distribution in the  midsection yet again resembles \blue{surface neuromast} patterns found in live fish \blue{(Figure~\ref{fig:fishNeuromasts}), and indicates that fish extremities may be ideal for detecting variations in shear stress.}

\blue{
\subsection{Optimal pressure gradient sensors}
\label{sec:swimmingFishPressGrad}

We now consider the optimal placement of pressure gradient sensors on the larva's body. These sensors are analogous to canal neuromasts found in live fish, which display markedly similar distribution patterns across a variety of fish species~\citep{Ristroph2015}. The canal is usually present in a continous line running from head to tail, and shows a high concentration of neuromasts in canal branches found in the head~\citep{Coombs1988,Ristroph2015}. We use a combination of the five distinct flow configurations described earlier, to determine the optimal arrangement of pressure gradient sensors by following the procedure described in section~\ref{sec:combined_shear}. The resulting utility curve and sensor distribution are shown in Figure~\ref{fig:combination_pressGrad}.
\begin{figure}
        \centering
        \begin{subfigure}[c]{\textwidth}
                \centering
                \includegraphics[width=0.8\textwidth]{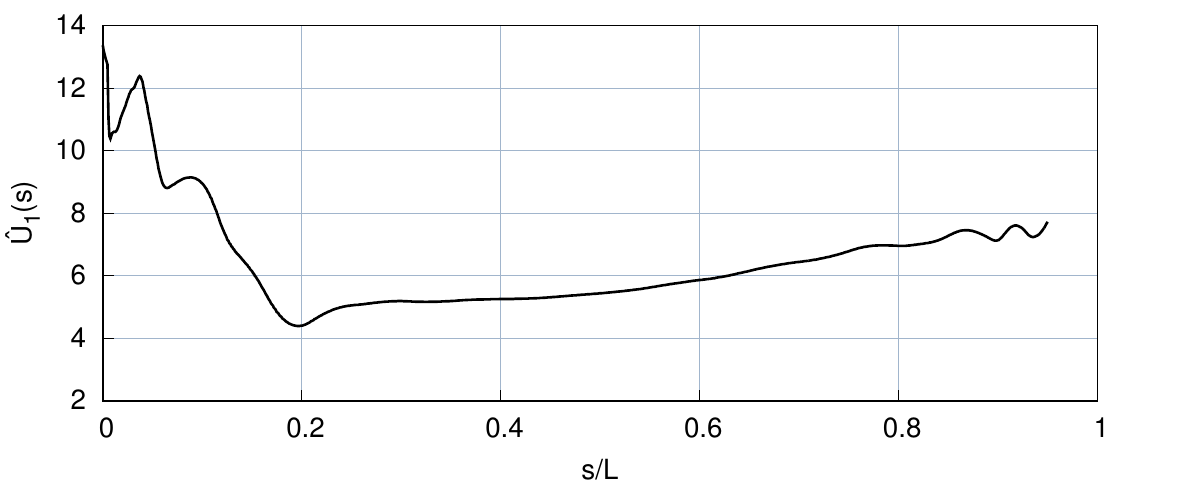}
                \subcaption{}
                \label{fig:utilityAllSum_pressGrad}
        \end{subfigure} \\
        \begin{subfigure}[c]{\textwidth}
                \centering
                \includegraphics[width=0.8\textwidth]{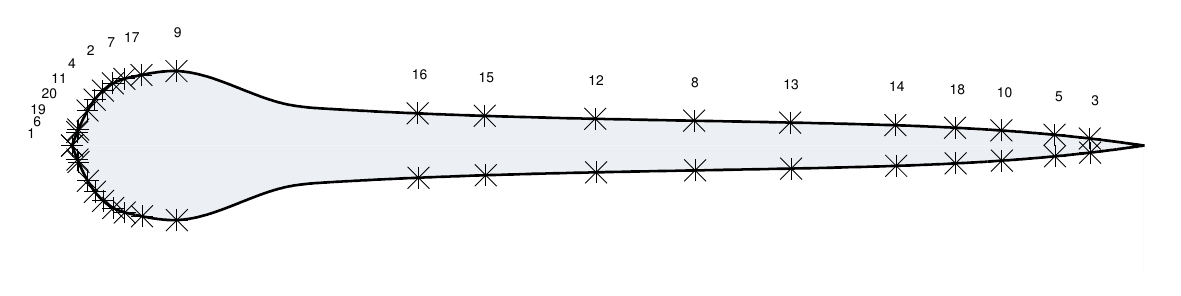}
                \subcaption{}
                \label{fig:sensorsAllSum_pressGrad}
        \end{subfigure}
		\caption{\blue{(\subref{fig:utilityAllSum_pressGrad}) Utility curve for the first \blue{pressure gradient} sensor, $\hat{U}_1(s)$, on a larva-shaped swimmer, using a combination of all five flow configurations described in the paper. (\subref{fig:sensorsAllSum_pressGrad}) Sequential placement of 20 \blue{pressure gradient} sensors along the body.}}
\label{fig:combination_pressGrad}
\end{figure}
The most notable difference between the arrangement of pressure gradient sensors (Figure~\ref{fig:sensorsAllSum_pressGrad}), and that of shear stress sensors (Figure~\ref{fig:sensorsAllSum}), is observed in the midsection of the body. We  find a consistent distribution of pressure gradient sensors in the midsection, which is not the case for shear stress sensors. Out of the 20 pressure gradient sensors placed, 10 are found clustered densely in the head ($s/L\le0.1$, which corresponds to high utility values in Figure~\ref{fig:utilityAllSum_pressGrad}), and the other 10 are spaced regularly throughout the body. This arrangement is similar to the neuromast distribution found in subsurface canals, which yet again suggests that this sensory structure may have evolved for detecting changes in pressure gradients with high accuracy. \refOne{In fact, the utility curve shown in Figure~\ref{fig:utilityAllSum_pressGrad} agrees qualitatively with the canal density reported by \citet{Ristroph2015}, especially for $s/L<0.2$. However, a direct comparison must be made with care, given that our simulations are two-dimensional, whereas the distributions reported by \citet{Ristroph2015} display significant three-dimensional branching in the head.} 
}


\subsection{Inference of disturbance-generating source}
\label{sec:inference}

Having determined the optimal distribution of sensors on the swimmer body, we now assess how effectively these arrangements can characterize the disturbance sources. For a given set of sensors $\sensors$, this involves estimating the probability that a particular \blue{sensor} measurement \blue{may} originate from different cylinder positions within the prior-region. 
For this, we consider the measurements $\Measured^{(GT)}$ at the sensor locations, generated from a single cylinder \blue{located at $\cylinderPos^{(GT)}$} (the superscript $GT$ denotes `ground-truth'). For a given sensor configuration $\sensors$, the measurements $\Measured^{(GT)}$ are computed using the prediction error model $\Measured^{(GT)} = \Signal( \cylinderPos^{(GT)}; \sensors) + \eps(\sensors)$, where $\Signal( \cylinderPos^{(GT)}; \sensors)$ is obtained by simulating the Navier-Stokes equations with an oscillating cylinder located at $\cylinderPos^{(GT)}$, and $\eps(\sensors)$ is a vector sampled from the Gaussian distribution $\mathcal{N}(0, \Sigma(\sensors))$. Assuming that the disturbance position $\cylinderPos^{(GT)}$ is unknown, the swimmer attempts to identify it by assigning probability values to all possible cylinder locations $\cylinderPos$ within the prior-region (i.e., by determining the posterior distribution $p(\cylinderPos | \Measured^{(GT)} ,\sensors)$). The highest probability value yields the best estimate for the cylinder position. This process is analogous to a fish attempting to localize the position of a predator or prey. Using the fact that the prior distribution of the disturbance location is uniform, 
the required posterior probability distribution of the disturbance location is proportional to the likelihood $p(\Measured \vert \cylinderPos^{(GT)}, \sensors)$ defined in Eq.~\ref{eq:likelihood}, where $\Measured$ are the measurements recorded along the swimmer body. 

\begin{figure}
	\centering
	\begin{subfigure}[c]{0.48\textwidth}
		\centering
		\includegraphics[width=\textwidth]{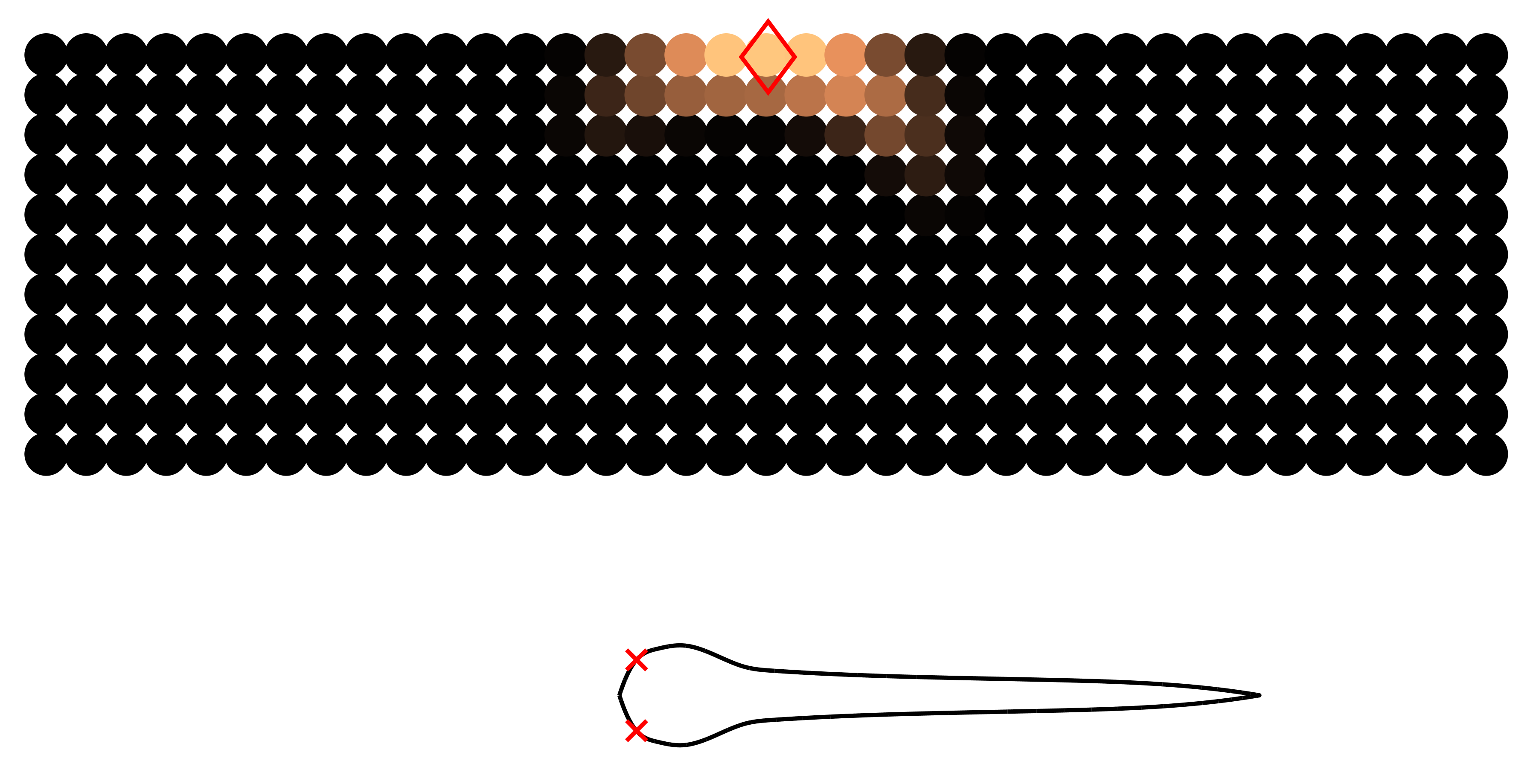}
		\subcaption{}
		\label{fig:1sensor_optimal}
	\end{subfigure}
	\begin{subfigure}[c]{0.48\textwidth}
		\centering
		\includegraphics[width=\textwidth]{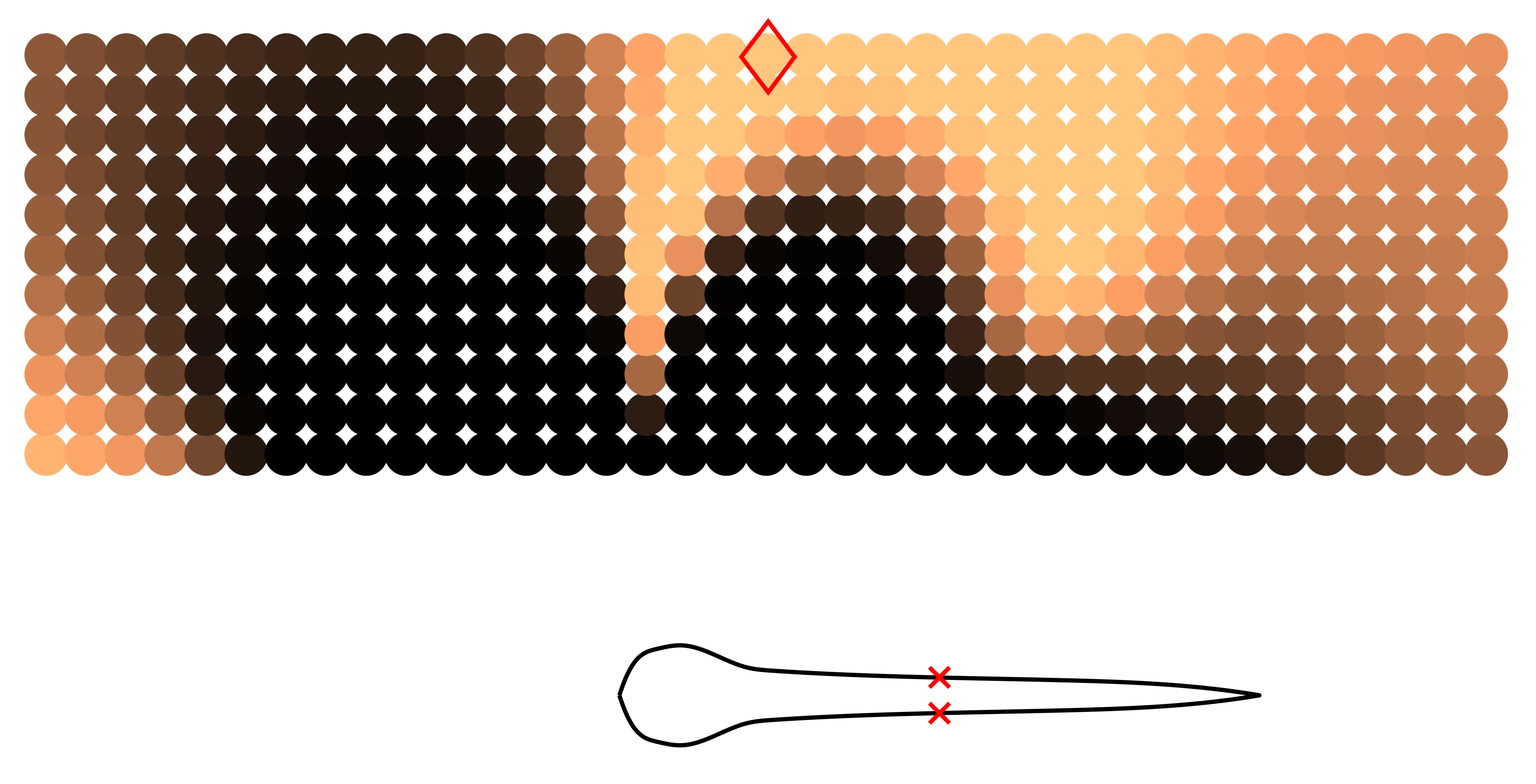}
		\subcaption{}
		\label{fig:1sensor_uniform}
	\end{subfigure} \\     
	\begin{subfigure}[c]{0.48\textwidth}
		\centering
		\includegraphics[width=\textwidth]{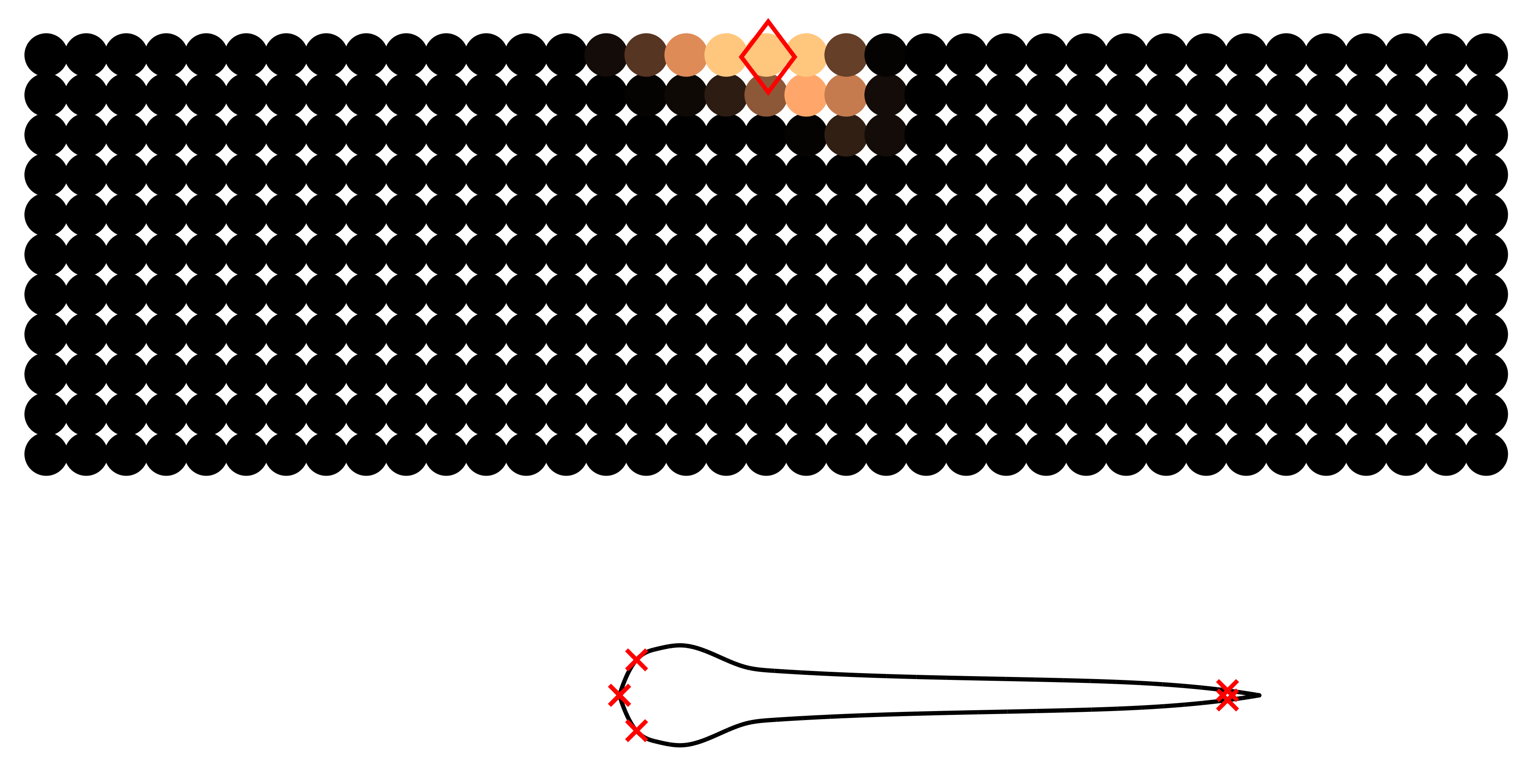}
		\subcaption{}
		\label{fig:3sensor_optimal}
	\end{subfigure}
	\begin{subfigure}[c]{0.48\textwidth}
		\centering
		\includegraphics[width=\textwidth]{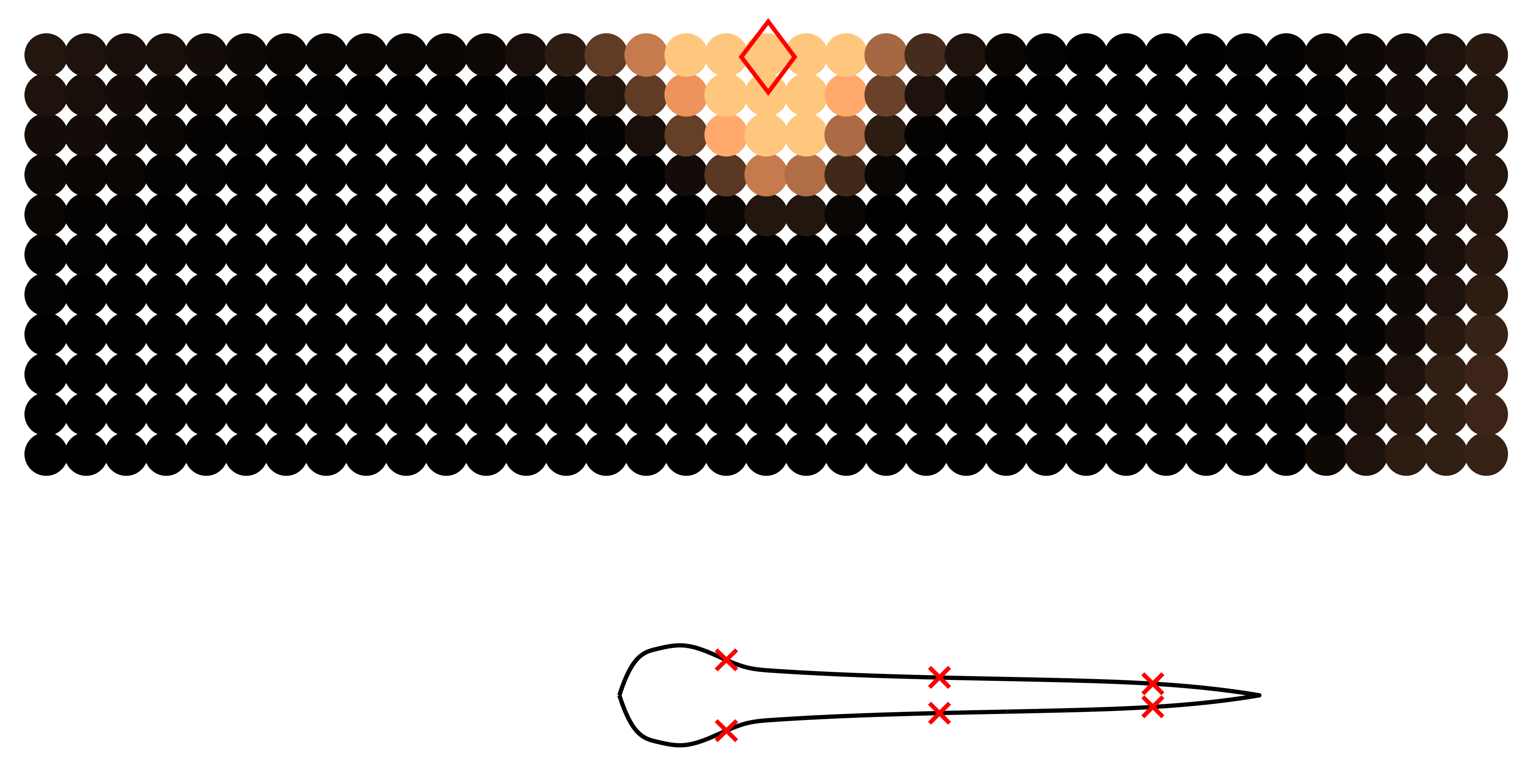}
		\subcaption{}
		\label{fig:3sensor_uniform}
	\end{subfigure} \\
	\begin{subfigure}[c]{0.48\textwidth}
		\centering
		\includegraphics[width=\textwidth]{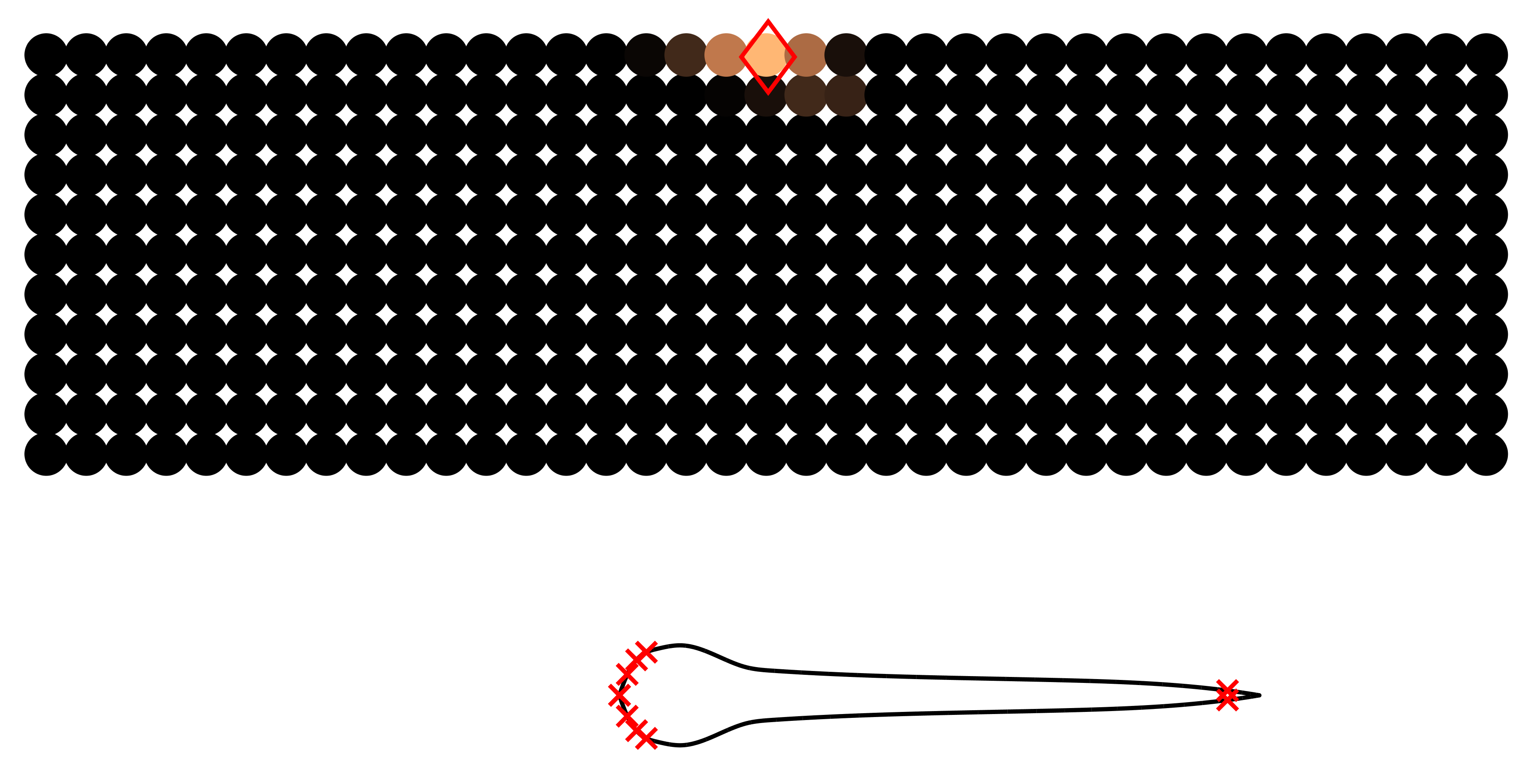}
		\subcaption{}
		\label{fig:5sensor_optimal}
	\end{subfigure}
	\begin{subfigure}[c]{0.48\textwidth}
		\centering
		\includegraphics[width=\textwidth]{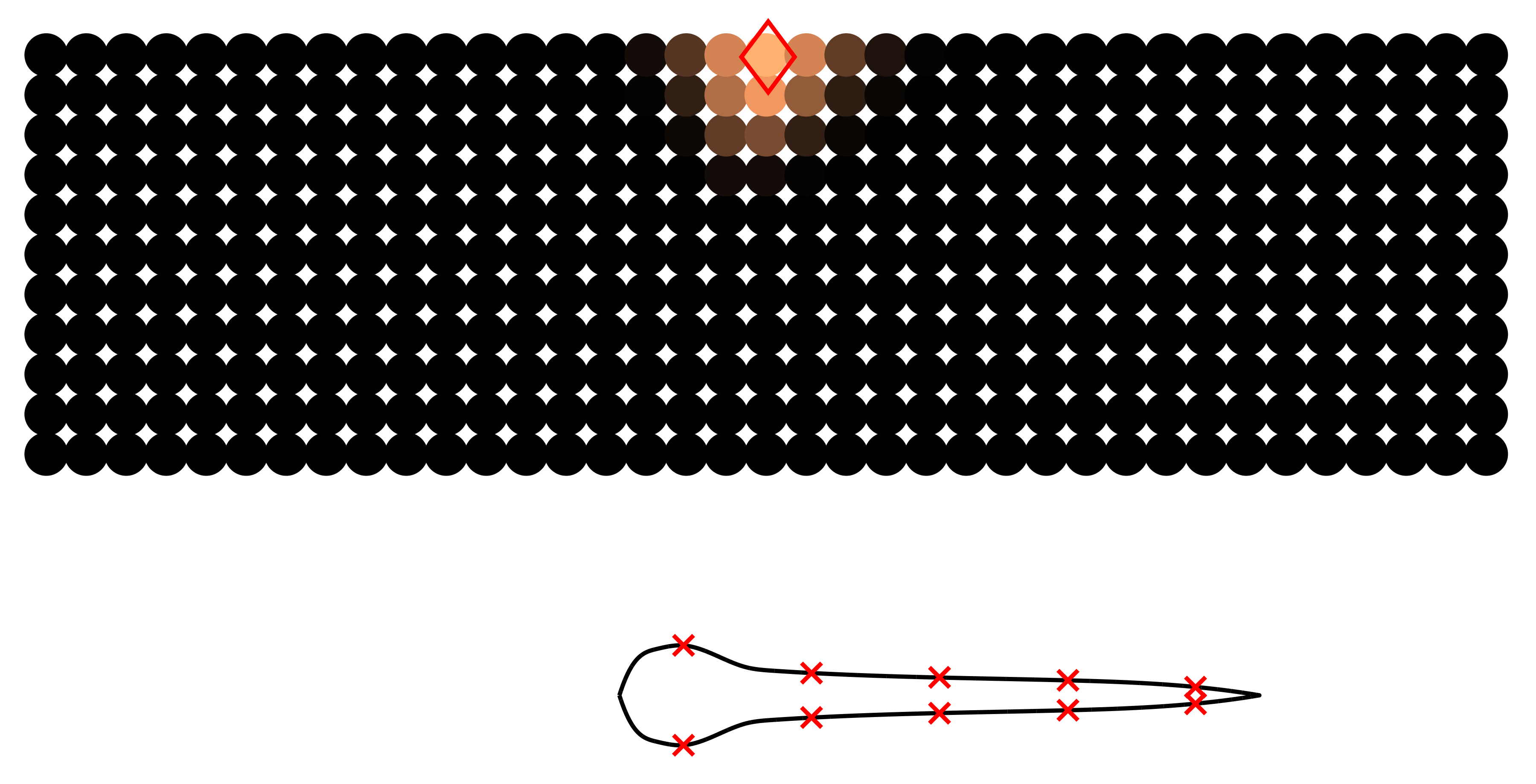}
		\subcaption{}
		\label{fig:5sensor_uniform}
	\end{subfigure}
	\caption{Plots showing the probability that the \refOne{signal being measured by a self-propelled swimmer} originates from a particular location in the prior-region. \refOne{Left column - swimmer using flow measurements from optimal sensor locations indicated on the body, right column - estimates made by a swimmer using suboptimal sensor distributions}. Brighter areas indicate regions of higher probability. The relevant sensor arrangement is shown using red `$\times$' symbols on the swimmers' bodies. The actual position of the signal-generating cylinder is marked with red diamonds. (\subref{fig:1sensor_optimal}, \subref{fig:3sensor_optimal}, \subref{fig:5sensor_optimal}) Probability distributions computed using measurements from 1, 3, and 5 optimal sensors on both sides of the body. (\subref{fig:1sensor_uniform}, \subref{fig:3sensor_uniform}, \subref{fig:5sensor_uniform}) Probability distributions for 1, 3, and 5 uniformly-distributed sensors on both sides.}
	\label{fig:inference}
\end{figure}

The resulting probability distribution for \blue{estimating the correct}  $\cylinderPos^{(GT)}$ is depicted in Figure~\ref{fig:inference}, with the \refOne{rows} showing results for an increasing number of sensors. \refOne{The left column shows probability-estimates from a self-propelled swimmer attempting to identify the position of an unknown disturbance-source, based on flow measurements from optimal sensor locations which are indicated on the body. The right column depicts estimates made by a swimmer using suboptimal sensor distributions. The probability distributions indicate that the swimmer on the left (using optimal sensor distributions) is able to provide a much more accurate estimate of the correct location of the disturbance-source.}
We observe that the un-informed placement of a single sensor in Figure~\ref{fig:1sensor_uniform} leads to a large spread in the probability distribution, making it difficult to locate the disturbance source accurately. In comparison, the first optimal sensor in Figure~\ref{fig:1sensor_optimal} yields a noticeably narrower spread, centered close to the correct position of the signal-generating cylinder (i.e., the ground-truth). The probability distributions in both cases become narrower with increasing number of sensors, making it easier to locate the disturbance source. \blue{In all cases, the optimal arrangement of sensors performs noticeably better than the uniform distribution for identifying the correct cylinder location.}

\section{Discussion}
\label{sec:discussion}

\refThree{While the present work attempts to represent realistic flow conditions, it is important to keep in mind its limitations and simplifying assumptions used in the simulations and data analyses. Most importantly, we note that flexural dynamics of the sensory structures are not accounted for in the current study. In actual fish, the extent to which the hair-like sensory structures are deflected by the flow can influence their sensing-effectiveness \citep{Hudspeth1977,Blake2006,VanTrump2008,Bleckmann2009}. For instance, beyond a certain locomotion speed, the superficial neuromasts may suffer from saturation effects during forward motion, when the hair cells are fully deflected and may no longer be able to detect external stimuli effectively. In addition to natural swimmers, such dynamics have been included in artificial lateral lines employed in robotic devices \citep{Fan2002}. Similarly, the canal neuromasts are located in recessed channels under the skin, which alter the flow experienced by the sensory structures considerably. This is compensated in real fish through the introduction of resonant hair cell structures \citep{Maoileidigh2012}, but is not accounted for in our simulations. All of these aspects may play an important role in determining the observed distribution of sensors on a fish's body, in addition to the fluid-flow induced by external stimuli. However, our two-dimensional simulations do not account for these factors, in an attempt to keep the complexity of the Navier-Stokes simulations to manageable levels. 

Furthermore, we note that detecting flows by a stationary swimmer in a quiescent fluid with external perturbations is not equivalent to flow measurements by moving swimmers; the latter suffer from separation effects of the boundary layer, which can influence the location of optimal sensors. Consequently, we must be careful when considering sensor-placement using the combined datasets, as done in Figures~\ref{fig:combination} and~\ref{fig:combination_pressGrad}. Inspecting the individual scenarios in Figures~\ref{fig:staticSensors}, \ref{fig:sensorsDcyl}, and \ref{fig:sensorsLarva}, we observe that the sensor distributions for the stationary and moving swimmers are not entirely dissimilar, which gives us confidence in using the combined dataset for sensor placement in Figures~\ref{fig:combination} and~\ref{fig:combination_pressGrad}.}

\refOne{In our approach, a sensor-pair is placed on the body surface symmetrically around the fish centerline. We anticipate that re-evaluating the simulations with a left-right flip, i.e., changing the orientation of the fish with respect to its approach to the cylinder, may lead to some differences in the exact sensor locations. However, once the fish is experiencing the cylinder's wake on its pressure and shear sensors, we expect that the overall sensor arrangement will remain unchanged, i.e., we will still observe a dense distribution of sensors in the head and the tail. We have found that the dominant factor in determining sensor placement is not the approach to the flow but rather the body-geometry and motion. This is evident when we compare sensor distributions across very dissimilar scenarios, i.e., the three different flow-configurations involving stationary fish, self-propelled fish, and rigid fish placed in the D-cylinder's wake (Figures~\ref{fig:staticSensors}, \ref{fig:sensorsLarva}, and \ref{fig:sensorsDcyl}).}

\refTwo{The present approach is computationally demanding as it requires conducting a large number of Direct Numerical Simulations (DNS) for the Navier-Stokes equations, each of which takes 4 to 7 hours to complete on a 12-core CPU node depending on the flow configuration. The computational cost is substantial, given that close to 400 distinct simulations have to be evaluated for each of the considered flow configurations, resulting in a total of approximately 3000 simulations. At the same time, we must emphasize that all DNS computations are performed once (off-line) and the sequential sensor-placement algorithm is rather inexpensive, with the computations taking on the order of 5 minutes using a single CPU core. 

Hence, it is imperative to store DNS data containing all possible sensor-measurements offline, and to process them for sequential sensor placement as required. We remark that if the same study were to be conducted in a three-dimensional setting using Direct Numerical Simulations, the computational cost of each simulation would increase by approximately three orders of magnitude, since the computational grid would increase in size from 4096 x 4096 cells for the present 2D cases, to approximately 4096 x 4096 x 1024 cells for the 3D cases. This would represent a significant increase in computational cost, especially considering the need to run approximately 3000 DNSs to examine all of the cases discussed in the present work. We remark that if, for 3D flows, we use instead approaches such as Large Eddy Simulations (LES) and Unsteady Reynolds Averaged Navier-Stokes (URANS) calculations we do not expect to have higher computational costs than the ones required herein for the 2D DNS. In closing, we argue that the availability of computational power and automation in the near future will make approaches such as the one presented herein amenable to further computational (and experimental) studies.}


\section{Conclusion}
\label{sec:conclusion}
We  have combined two-dimensional Navier-Stokes simulations with Bayesian optimal experimental design, to identify the best arrangement of sensory structures on self-propelled swimmers' bodies. The study is inspired by the particular distribution of flow-sensing mechanoreceptors found in many fish species, referred to as the lateral-line organ, where a large number of sensory structures are located in the head. 

We optimize sensor arrangements on two different swimmer shapes under the influence of various sources of disturbance. We find  optimal  arrangements that resemble those found in fish bodies, suggesting that such arrangements may allow them to gather information from their surroundings more effectively than other layouts. We demonstrate that the optimal configuration of these sensors depends on the body shape and the type of disturbance being perceived. This is explored using a variety of simulations involving both static and swimming configurations, using distinct body profiles resembling fish larvae and adults, and using disturbances generated by oscillating \blue{cylinders,}  rotating cylinders\blue{,} and by D-shaped half cylinders. Despite certain differences that exist in sensor distributions among the various cases considered, there is a marked tendency for a large number of \blue{shear stress} sensors to be located in the head \blue{and the tail} of the swimmer, with \blue{virtually no sensors found}  in the midsection. \blue{In the case of pressure gradient sensors, we observe a high density of sensors placed in the head, followed by regularly spaced distribution along the entire body.}  \blue{These observations} closely reflect the structure of the sensory organ in live fish.

To assess the effectiveness of the sensor placement algorithm, we compare the performance of optimal arrangements to that of un-informed uniform sensor distributions. The results confirm that optimal distribution patterns lead to more accurate identification of external disturbances, which suggests that these distinctive distributions may allow fish to assimilate maximum information from their surroundings using the fewest number of neuromasts. We believe that the present work is a positive step towards understanding mechanosensing in fish, and we hope that the proposed  methodology  can assist in the development of optimal sensory-layouts for engineered swimmers. 

\section{Acknowledgments}
\blue{This work was supported by European Research Council Advanced Investigator Award 341117, and utilized computational resources granted by the Swiss National Supercomputing Centre (CSCS) under project IDs `s658' and `ch7'. We thank Panagiotis E. Hadjidoukas for assistance with the TORC interface for running the flow simulations.}
\bibliography{optimalSensor}
\bibliographystyle{jfm}

\clearpage
\pagebreak

\appendix

\section{Shape parametrisation and swimming kinematics}
\label{app:shapeKinematics}
The adult shape used in the simulations is modelled using three piecewise polynomials,
\begin{align}
h(s) = \begin{cases}
\sqrt{2 s h_{head} - s^2} &, 0\le s<s_{head} \\
	h_{head} - (h_{head} - h_{tail})\left(\dfrac{s-s_{head}}{s_{tail} - s_{head}}\right)^2 &, s_{head}\le s<s_{tail}\\
h_{parabola} &, s_{tail}\le s \le L \PERIOD
\end{cases}
\label{eq:adultShape}
\end{align}
Here, $s$ denotes the curvilinear coordinate running from the head to the tail along the body midline, and $h$ is the half-width of the body. The constant parameters that determine the final shape are: $s_{head} = 0.04L$, $s_{tail} = 0.95L$, $h_{head} = 0.04L$, $h_{tail} = 0.01L$. The parabolic section that defines the smooth tail is computed as follows:
\begin{subequations}
\begin{align}
a &= h_{tail}\\
b &= \dfrac{h_{tail} - h_{head}}{s_{tail} - s_{head}} \\
c &= \dfrac{-b(L - s_{tail}) -  h_{tail}}{(L-s_{tail})^2} \\
h_{parabola} &= a + b(s-s_{tail}) + c(s-s_{tail})^2 \PERIOD
\end{align}
\end{subequations}

The larva shape used in the simulations is based on silhouettes of zebrafish extracted from experiments\footnote{The authors thank the Engert Lab at Harvard University for providing the experimental images.}. The segmented shape is parametrized using a natural cubic spline comprised of 7 piecewise sections, with the following knots and polynomial coefficients: 
\begin{subequations}
\label{eq:larvaShape}
\begin{align}
\left(s_0, \cdots\, s_i, \cdots, s_7\right)/L  &=  \left(0.0, 0.018, 0.058, 0.098, 0.198, 0.238, 0.698, 1.0\right) \\
a_{i,j} &=
\begin{pmatrix*}[r]
                            &0.000000      &3.152700        &-44.18100      &145.49 \\
                            &0.043282      &1.703600        &-36.32400      &300.38 \\
                            &0.072530      &0.239460        &-0.278770      &-54.342 \\
                            &0.078185      &-0.043688       &-6.799800      &38.155 \\
                            &0.043973      &-0.258990       &4.646700       &-37.601 \\
                            &0.038641      &-0.067740       &0.134620       &-0.12819 \\
                            &0.023488      &-0.025271       &-0.042293      &-0.43563
\end{pmatrix*}
\end{align}
\label{eq:splineCoeffs}
\end{subequations}
Here, subscript $i\in\left[1,7\right]$ denotes the polynomial segment between knots $s_{i-1}$ and $s_i$. The corresponding cubic polynomial describing the body half-width in each of the 7 sections is given by
\begin{equation}
h_i(s) = a_{i,1} + a_{i,2} \left(s-s_{i-1}\right) + a_{i,3} \left(s-s_{i-1}\right)^2 + a_{i,4} \left(s-s_{i-1}\right) ^3 \PERIOD
\end{equation}

The swimmers propel themselves by imposing a sinusoidal wave travelling along the body. This wave is specified by the spatially- and temporally-varying curvature $\kappa(s,t)$ of the mid-line, 
\begin{equation}
\kappa(s,t) = A(s) \sin\left(\dfrac{2\pi t}{T} - \dfrac{2\pi s}{L} + \phi\right) \COMMA
\label{eq:kappaSin}
\end{equation} 
where $A(s)$ is described using a natural cubic spline, with 6 control points located at $(s_i, A_i) = ({ (0, 0.82), (0.15L, 1.47), (0.4L, 2.57), (0.65L, 3.75), (0.9L, 5.09), (1L, 5.7) )} $. Here, $s$ denotes the curvilinear coordinate running from the head to the tail along the body mid-line. The time period of body-undulation is set to $T=0.4$, and the phase difference $\phi$ is initially set to 0. The time-varying curvature is imposed as a function of the arc-length $s$, ensuring that the swimmer's length remains constant during deformation. The lateral and longitudinal coordinates (as well as the deformation velocities) of the mid-line points are recovered from the curvature using the Frenet-Serret formulas. The travelling wave form yields steadily-swimming fish. However, several species employ an intermittent swimming mode referred to as `burst-and-coast' swimming \citep{Weihs1974}. 
The burst and coast mode may not only be energetically efficient, but it also allows fish to better accumulate information about external disturbances, by stabilizing the sensory fields \citep{Kramer2001}. The burst-and-coast strategy is observed in blind cave fish, where they accelerate and glide past unfamiliar objects and obstacles repeatedly \citep{vonCampenhausen1981}. This allows them to form a `hydrodynamic image' of their surroundings, by perceiving reflections of their self-generated motion. Intermittent swimming has also been shown to be critical for avoiding collisions when approaching a wall \citep{Windsor2008}. The gliding phase during burst-coast swimming may minimize self-generated `noise' in the boundary layer on the body, thereby allowing signals of external origin to permeate through to the neuromasts. Modeling the the burst-coast motion for the swimmers involves multiplying the curvature amplitude $A(s)$ with a smoothly varying piecewise function $f(t)$, as described in  \citet{Verma2017CEC}:
\begin{equation}
f(t)=\left\{\begin{array}{ll} 1 & t \in \Delta t_{steady}\\
1-3\lambda_{coast}^2+2\lambda_{coast}^3 & t \in \Delta t_{decel}\\
0 & t \in \Delta t_{coast} \\
3\lambda_{burst}^2-2\lambda_{burst}^3 & t \in \Delta t_{accel} \PERIOD
\end{array}\right. 
\end{equation}
Here, $\lambda_{coast},\lambda_{burst}\in[0,1]$ are ramp functions increasing linearly from 0 to 1 within the intervals $\Delta  t_{decel}$ and $ \Delta t_{accel}$. The 4 time-intervals describing the burst-coast phases are set to $\Delta t_{steady} = 1.5T$, $\Delta t_{decel} = 0.375T$, $\Delta t_{coast} = 1.0T$, and $\Delta t_{accel} = 0.375T$. At the end of each coasting phase, the phase angle in Eq.~\ref{eq:kappaSin} is updated to $\phi = \left(\pi - 2\pi t_{SA}/T\right)$, where $t_{SA}$ denotes the time  at the start of the acceleration phase (or, equivalently, at the end of the coasting phase). This introduces mirror symmetry between tail-beats from one burst-coast cycle to another, and allows the fish to swim on a relatively straight trajectory.

\section{Derivation of the utility estimator}
\label{sec:utilityDerivation}

%
%
%
%

By applying Bayes' theorem Eq.~\ref{eq:utility}  can be written equivalently as,
\begin{equation}\label{eq:estimator:a}
U(\sensors) =  \int_{\YSPACE}  \int_{\PSPACE}  \;  \ln \frac{p( \Measured | \cylinderPos,\sensors)}{p(\Measured | \sensors)} \; p(\cylinderPos) \; p(\Measured | \cylinderPos,\sensors)  \dd{\cylinderPos} \dd{\Measured}  \COMMA
\end{equation}
where we have used the assumption that $p(\cylinderPos)=p(\cylinderPos | \sensors)$.
We approximate the integral over $\PSPACE$ with a quadrature rule on the points $\cylinderPos^{(i)}$ and weights $w_i$ for $i=1,\ldots,N_{\cylinderPos}$. Then Eq.~\ref{eq:estimator:a} is approximated by,
\begin{equation}
U(\sensors) \approx   \sum_{i=1}^{N_{\cylinderPos}} w_i  \; p(\cylinderPos^{(i)})  \,  \int_{\mathcal{Y}} \,   \ln \frac{p( \Measured | \cylinderPos^{(i)},\sensors)}{p(\Measured | \sensors)}  \; p(\Measured | \cylinderPos^{(i)},\sensors)   \dd{\Measured} \PERIOD
\end{equation}
For each $i$ in the summation, the integral over $\mathcal{Y}$ is approximated by Monte Carlo integration using $N_{\DATA}$ points $\{\DATA^{(i,j)} \}_{j=1}^{N_{\DATA}}$ from $p_{\Measured}( \,\cdot\, | \cylinderPos^{(i)},\sensors)$, leading to 
\begin{equation}
U(\sensors) \approx   \sum_{i=1}^{N_{\cylinderPos}}  \sum_{j=1}^{N_{\DATA}}   \frac{w_i  \; p(\cylinderPos^{(i)})}{N_{\DATA}}  \, \Big [  \ln {p( \Measured^{(i,j)} | \cylinderPos^{(i)},\sensors)}     - \ln{p(\Measured^{(i,j)} | \sensors)} \Big ] \PERIOD
\label{B3}
\end{equation}
The quantity $p( \Measured^{(i,j)} | \sensors)$ is approximated with the same quadrature rule,
\begin{equation}
\begin{split}
p( \Measured^{(i,j)} | \sensors)   &=   \int_{\PSPACE} \, p( \Measured^{(i,j)} | \cylinderPos,\sensors) \; p(\cylinderPos) \dd{\cylinderPos} \\
& \approx \sum_{k=1}^{N_{\cylinderPos}} w_k  \;  p( \Measured^{(i,j)} | \cylinderPos^{(k)},\sensors) \; p(\cylinderPos^{(k)}) \PERIOD
\end{split}
\label{B4}
\end{equation}

Finally, substituting Eq. \ref{B4} into Eq. \ref{B3}, the estimator for $U(\DES)$ is given by Eq. \ref{eq:estimator}. 


\section{Utility of combined experiments} \label{sec:app:derivation}

We assume that the measurements $\DATA_1$ and $\DATA_2$ are independent conditioned on $\DES$. We want to show that the utility function using both experiments $U_{1,2}$ is equal to the sum of the individual utility functions, $U_1$ and $U_2$.
\begin{equation}\label{ap:eq:ut:tot}
\begin{split}
U_{1,2}(\DES) &= 
 \int_{\mathcal{Y}_1}\int_{\mathcal{Y}_2}  \int_{\Theta}  \;  \ln \frac{p( \DATA_1,\DATA_2 | \PAR,\DES)}{p(\DATA_1,\DATA_2 | \DES)} \; p(\PAR) \; p(\DATA_1,\DATA_2 | \PAR,\DES)  \dd{\PAR} \dd{\DATA_1} \dd{\DATA_2}     \\
&= \int_{\mathcal{Y}_1}\int_{\mathcal{Y}_2}  \int_{\Theta}  \;  \ln \frac{  p( \DATA_1 | \PAR,\DES) p( \DATA_2 | \PAR,\DES)  }{p(\DATA_1,\DATA_2 | \DES)} \; p(\PAR) \; p(\DATA_1,\DATA_2 | \PAR,\DES)  \dd{\PAR} \dd{\DATA_1} \dd{\DATA_2}  \\   
&= \int_{\mathcal{Y}_1}  \int_{\Theta}  \;  \ln  p( \DATA_1 | \PAR,\DES)  \, p(\PAR) \ p(\DATA_1 | \PAR,\DES)  \dd{\PAR} \dd{\DATA_1} \\
&  \hspace{40pt} + \int_{\mathcal{Y}_2}  \int_{\Theta}  \;  \ln  p( \DATA_2 | \PAR,\DES)  \, p(\PAR) \ p(\DATA_2 | \PAR,\DES)  \dd{\PAR} \dd{\DATA_2} - h(\DATA_1,\DATA_2 | \DES) \COMMA
\end{split}
\end{equation}
where 
\begin{equation}\label{eq:C2}
h(\DATA_1,\DATA_2 | \DES) =  \int_{\mathcal{Y}_1}\int_{\mathcal{Y}_2}  \int_{\Theta}  \;  \ln p(\DATA_1,\DATA_2 | \DES) \; p(\PAR) \; p(\DATA_1,\DATA_2 | \PAR,\DES)  \dd{\PAR} \dd{\DATA_1} \dd{\DATA_2}  \COMMA
\end{equation}
Using Eq. \ref{eq:estimator:a}, it is easy to check that
\begin{equation}\label{ap:eq:ut:i}
U_i(\DES) = \int_{\mathcal{Y}_i}  \int_{\Theta}  \;  \ln  p( \DATA_i | \PAR,\DES)  \, p(\PAR) \ p(\DATA_i | \PAR,\DES)  \dd{\PAR} \dd{\DATA_i} - h(\DATA_i | \DES)  \COMMA 
\end{equation}
for $i=1,2$, where $h(\DATA_i | \DES)$ is given by
\begin{equation}\label{eq:C4}
h(\DATA_i | \DES) =  \int_{\YSPACE}  \int_{\PSPACE}  \;  \ln p(\DATA_i | \sensors) \; p(\PAR ) \; p(\DATA_i | \PAR ,\sensors)  \dd{\PAR} \dd{\DATA_i}  \COMMA
\end{equation}
Substituting Eq.~\ref{ap:eq:ut:i} into Eq.~\ref{ap:eq:ut:tot},
\begin{equation}
U_{1,2}(\DES) = U_1(\DES) + U_2(\DES)  - \Big[ h(\DATA_1,\DATA_2 | \DES) - h(\DATA_1| \DES) - h(\DATA_2 | \DES) \Big ] \PERIOD
\end{equation}
Using Eq. \ref{eq:C2} and Eq. \ref{eq:C4}, along with the fact that $\DATA_1$ is independent of $\DATA_2$  conditioned on $\DES$, the term inside the brackets is equal to zero, leading to 
\begin{equation}
U_{1,2}(\DES) = U_1(\DES) + U_2(\DES)  \PERIOD
\end{equation}
The assumption of conditional independency is valid in our case since knowledge of the measurements of one experiment $\DATA_1$ and the location of the sensors $\DES$ does not provide any additional information for the measurements of the other experiment compared to the case of knowledge only of the sensor location. Equivalently, $p(\DATA_2 | \DATA_1, \DES) =  p(\DATA_2 | \DES)$.

\end{document}